\documentclass[runningheads,a4paper,final]{llncs}
\usepackage[T1]{fontenc}

\usepackage[obeyFinal]{todonotes}
\usepackage{etoolbox,xparse}
\newrobustcmd{\dn}[1]{\todo[color=cyan]{#1}}
\newrobustcmd{\dni}[1]{\todo[inline,color=cyan]{#1}}
\newrobustcmd{\bm}[1]{\todo{#1}}

\usepackage{xspace}
\usepackage{graphicx}
\usepackage{mathtools}
\usepackage{amsbsy,amssymb}
\usepackage{amsfonts}
\usepackage{amstext}

\usepackage{tikz}
\usetikzlibrary{shapes,automata,arrows,decorations.pathmorphing,positioning}
\tikzstyle{every state}=[minimum size=2em]
\tikzset{every picture/.style={>=latex,auto,node distance=2cm, every
		loop/.style={looseness=6}, initial text={}, inner sep=1mm,
		loopright/.style={loop,looseness=6,out=35, in=-35},
		loopleft/.style={loop,looseness=6,out=145, in=215},
		loopabove/.style={loop,looseness=6,out=125, in=55},
		loopbelow/.style={loop,looseness=6,out=-125, in=-55}}}

\newcommand\N{\mathbb N}

\newcommand\FO{\ensuremath{\mathsf{FO}}\xspace}
\newcommand\twoWFO{\ensuremath{\mathsf{WFO}}\xspace}
\newcommand\WFO{\twoWFO}
\newcommand\K{\mathsf{K}\xspace}

\newcommand\lrWFO{\ensuremath{\mathsf{WFO}^{\rightarrow}}\xspace}

\newcommand\rlWFO{\ensuremath{\mathsf{WFO}^{\leftarrow}}\xspace}
\newcommand\RoneWFO{\ensuremath{\mathsf{rWFO}^{\rightarrow}}\xspace}

\newcommand\A{\mathcal A}
\newcommand\B{\mathcal B}

\newcommand\zero{\ensuremath{\mathbf 0}\xspace}
\newcommand{\one}{\ensuremath{\mathbf 1}\xspace}

\newcommand\NWAr[1]{\ensuremath{{#1}\text{-}\mathsf{NWA}}\xspace}
\newcommand\NWA{\ensuremath{\mathsf{NWA}}\xspace}
\newcommand\lrNWAr[1]{\ensuremath{{#1}\text{-}\mathsf{NWA}^{\rightarrow}}\xspace}
\newcommand\rlNWAr[1]{\ensuremath{{#1}\text{-}\mathsf{NWA}^{\leftarrow}}\xspace}
\newcommand\lrNWA{\ensuremath{\mathsf{NWA}^{\rightarrow}}\xspace}
\newcommand\rlNWA{\ensuremath{\mathsf{NWA}^{\leftarrow}}\xspace}

\newrobustcmd{\swNWA}{\ensuremath{\mathsf{swNWA}}\xspace}
\newrobustcmd{\swNWAr}[1]{{#1}\text{-}\mathsf{swNWA}}

\newcommand{\aggr}{\mathsf{aggr}}

\newcommand\Wexists[1]{\Sigma_{#1}}
\newcommand\WFOrallLR[1]{\Pi_{#1}}
\newcommand\WFOrallRL[1]{\Pi^{-1}_{#1}}

\newrobustcmd{\twonwA}{\NWA}
\newrobustcmd{\swnwA}{\mathsf{swNWA}}

\newrobustcmd{\sem}[1]{\{\!| #1 |\!\}}
\newrobustcmd{\csem}[1]{[\![#1]\!]}
\newrobustcmd{\mult}[1]{\{\!\!\{#1\}\!\!\}}
\newrobustcmd{\word}[1]{\vartriangleright\!\! #1 \!\!\vartriangleleft}
\newrobustcmd{\trans}{\mathsf{Tr}}
\newrobustcmd{\init}{\mathsf{Init}}
\newrobustcmd{\fin}{\mathsf{Fin}}
\newrobustcmd{\dem}[1]{\mathsf{dem}^{#1}}
\newrobustcmd{\lift}[1]{\mathsf{lift}(#1)}
\newrobustcmd{\wt}{\mathsf{wt}}
\newrobustcmd{\tup}[1]{\langle #1 \rangle}
\newrobustcmd{\bh}[2]{\mathsf{bh}^{#1}_{#2}}
\newrobustcmd{\lmark}{{\vartriangleright}}
\newrobustcmd{\rmark}{{\vartriangleleft}}
\newrobustcmd{\trmon}[1]{\mathsf{TrMon}(#1)}
\newrobustcmd{\sw}[1]{{}^{\mathsf{sw}\!}#1}

\renewcommand{\epsilon}{\varepsilon}

\usepackage{hyperref}

\begin{document}
	
	\title{An Automata Theoretic Characterization of Weighted First-Order Logic\thanks{We thank the reviewers that helped greatly improving the readability of this article. The work was partially done during an internship of the first author at Aix-Marseille Université, partially funded by CNRS IRL 2000 ReLaX.}} 
	
	\author{Dhruv Nevatia\inst{1}\orcidID{0009-0008-0845-6754} 
		\\ 
		\and
		Benjamin Monmege\inst{2}\orcidID{0000-0002-4717-9955}
	}
	
	\institute{ETH Zurich,
		Switzerland
		\email{dhruv.nevatia@inf.ethz.ch}\and 
		Aix Marseille Univ, LIS, CNRS, Marseille,
		France
		\email{benjamin.monmege@univ-amu.fr}}
	
	\authorrunning{Dhruv N. and B. Monmege}
	
	\maketitle
	
	\begin{abstract}
		Since the 1970s with the work of McNaughton, Papert and Schützenberger \cite{Sch65,McNPap71}, a regular language is known to be definable in the first-order logic if and only if its syntactic monoid is aperiodic. This algebraic characterisation of a fundamental logical fragment has been extended in the quantitative case by Droste and Gastin \cite{droste2019aperiodic}, dealing with polynomially ambiguous weighted automata and a restricted fragment of weighted first-order logic. In the quantitative setting, the full weighted first-order logic (without the restriction that Droste and Gastin use, about the quantifier alternation) is more powerful than weighted automata, and extensions of the automata with two-way navigation, and pebbles or nested capabilities have been introduced to deal with it \cite{BolGas14,GasMon14}. In this work, we characterise the fragment of these extended weighted automata that recognise exactly the full weighted first-order logic, under the condition that automata are polynomially ambiguous.
	\end{abstract}
	
	\section{Introduction}

	Early works by McNaughton, Papert and Schützenberger \cite{Sch65,McNPap71} have enabled an automata-theoretic characterisation of first-order logic over finite words: a regular language is definable in the first-order logic if and only if its syntactic monoid is aperiodic. From the minimal automaton recognising the language, we can compute its syntactic monoid and check aperiodicity to conclude. Moreover, from the aperiodic minimal automaton, we can deduce a first-order formula equivalent to it. 
	
	More recently, Droste and Gastin \cite{droste2019aperiodic} have extended this result to deal with quantitative extensions of the first-order logic and automata. These quantitative extensions find their origin in the works of Schützenberger \cite{Sch61} that investigated \emph{weighted automata}, and their expressive power in terms of \emph{(formal power) series} that are mappings from finite words to weights of a \textit{semiring}. Weighted automata were originally thought as finite state automata where each transition (as well as initial and final states) are equipped with weights of a semiring. Along a run, weights are combined by the multiplication of the semiring, while non-determinism is resolved by considering the sum of the weights of all accepting runs over a word. By changing the semiring we consider, weights can model cost, rewards, energy or probabilities in a unified way: see \cite{DroKui09}. Many extensions have then been considered, by allowing for more structures than words (infinite words~\cite{DroRah06}, trees~\cite{DroVog11}, nested words~\cite{DroPib12,BolGas13}) and more weights than semirings (valuation monoids~\cite{DroMei12}, multioperator monoids~\cite{FulStu12}). 
	
	In order to describe the series describable by weighted automata in a more readable way, it might be useful to have more high-level ways of description, like weighted logics based on monadic second order (MSO) features, introduced by Droste and Gastin \cite{DroGas07}. Based on the seminal result by Büchi, Elgot, and Trakhtenbrot~\cite{Buc60,Elg61,Tra61}, they have explored a weighted extension of MSO logic on finite words and semirings: the semantics of disjunction and existential quantification are based on the sum of the semiring, while the ones of conjunction and universal quantification are based on the product. The appropriate restriction on the logic was found in order to obtain the exact same expressivity as weighted automata: a restriction is needed for combinatorial reasons, certain operators of the logic being able to generate series growing too quickly with respect to the length of the input word. In particular, universal quantifications must be used only once over very basic formulas, and conjunction is not allowed. Once again, this seminal result relating weighted automata and weighted logics has been extended in many ways: on trees 
	\cite{DroVog11}, on nested words \cite{DroPib12,BolGas13}, with weights valuation monoids \cite{DroMei12}, to cite only a very~few. 
	
	In \cite{GasMon18}, the semantics of weighted automata and weighted MSO logic has been revisited in a uniform way allowing one to obtain many previous results in a simplified way. First, an
	\textit{abstract semantics} is defined, mapping each word to a multiset of sequences of weights (one sequence per accepting run): this abstract semantics does not depend on the weight structure, since no actual computation is made. The abstract semantics can then be aggregated into a single output weight by an ad-hoc operator: we call this the \textit{concrete semantics}. Methodologically speaking, showing that two models have equal abstract semantics is sufficient (but not necessary in general) to show that they have equivalent concrete semantics.
	
	In \cite{droste2019aperiodic}, Droste and Gastin consider the first-order fragment \WFO of the weighted MSO logic, with the same kind of restrictions as the one explored for weighted MSO logic to recover the same expressive power as weighted automata. Under this restriction, they show that the logic \WFO is expressively equivalent to weighted automata that are \textit{aperiodic} (defined similarly as in the unweighted setting) and \textit{polynomially ambiguous}. Moreover, the proof is constructive and works for the abstract semantics (and thus for any concrete semantics).
	
	In order to express more properties than the restricted logics (\WFO and weighted MSO), weighted automata with two-way navigation and pebbles or nested capabilities have been introduced~\cite{BolGas14,GasMon14}, with an equivalent logic based on an extension of \WFO with some limited transitive closure operators. As noted in \cite{droste2019aperiodic} by Droste and Gastin, this is thus natural to ask what the models of two-way nested/pebble weighted automata are that recognise exactly the full \WFO logic. In this work, we answer this question: the series recognised by \WFO logic can be obtained from two-way nested/pebble weighted automata that are aperiodic and polynomially unambiguous. This generalises the results of \cite{droste2019aperiodic} only working for a small fragment of \WFO, and one-way (non-nested) weighted automata, but the condition has a similar flavour. The aperiodicity condition on two-way automata models has been explored in \cite{carton2021aperiodic}. Our proof is constructive and goes through a special case of two-way automata that are called \textit{sweeping} where every change of direction happens on the border of the input word (and not in the middle). This allows us to more easily reuse the work by Droste and Gastin, which only works for one-way models. 
	
	After defining the weighted first-order logic we study in Section~\ref{sec:logic}, and the nested two-way weighted automata in Section~\ref{sec:NWA}, we prove the equivalence between the various formalisms in subsequent sections: the translation from the logic to sweeping nested weighted automata is done in Section~\ref{sec:logic2aut}; sweeping nested weighted automata are translated back in the logic in Section~\ref{sec:sweeping2logic}. The most difficult part of the proof is the translation from two-way nested weighted automata to sweeping nested weighted automata: this does not hold if we do not have nesting mechanisms, and this translation thus raises the number of nesting necessary in the model.
	
	\section{Weighted First-Order Logic}\label{sec:logic}
	
	In this section, we introduce the weighted first-order
	logic whose power we will characterise in the following with respect to some automata model. The logic used in \cite{droste2019aperiodic} is a fragment of this logic where nesting of operations is limited to be as expressive as weighted automata.
	
	\begin{definition}
		For a set $\K$ of weights and an alphabet $A$, we let $\WFO(\K,A)$ be the logic defined by the following grammar: 
		\begin{align*}
			\varphi & \Coloneqq\top\mid P_{a}(x)\mid x\leq y\mid\neg\varphi\mid\varphi\wedge\varphi\mid\forall x\, \varphi & (\FO)\\
			\Phi & \Coloneqq\zero \mid\one \mid k \mid \varphi?\Phi:\Phi\mid\Phi+\Phi\mid\Phi\cdot\Phi\mid\Wexists{x}\Phi\mid \WFOrallLR{x}\Phi\mid\WFOrallRL{x}\Phi & (\WFO)
		\end{align*}
		where $a\in A$, $k\in\K$ and $x,y$ are first order variables.
	\end{definition}
	
	Formulas $\varphi$ stand for the classical (Boolean) first-order logic over words on the alphabet~$A$. Their semantics is defined classically over words $u=u_1u_2\cdots u_n\in A^{*}$ and valuations $\sigma\colon \mathcal{V}\rightarrow\{1,2,\ldots,n\}$ of the free variables $\mathcal V$ of the formula, letting $u, \sigma\models \varphi$ when the formula is satisfied by the word and the valuation. 
	
	Formulas $\Phi$ are weighted formulas that intuitively associate a weight with each word and valuation of free variables. As described in \cite{GasMon14}, the semantics is defined in two steps: first we give an \emph{abstract} semantics associating with each word and valuation a multiset of sequences of weights in $\K$; then we may define a \emph{concrete} semantics by describing how to fuse the multiset of sequences into a single weight. This differs from the classical semantics that directly compute the concrete semantics, but for our later proofs the other equivalent definition is much easier to manipulate. 
	
	Let $u\in A^{*}$
	be a word and $\sigma\colon\mathcal{V}\rightarrow\{1,2,\ldots,n\}$ be a
	valuation where $\mathcal{V}$ is a set of variables. The abstract semantics
	of a $\WFO$ formula $\phi$, with $\mathcal{V}$ as free variables, is denoted by $\sem{\phi}_{\mathcal V}(u, \sigma)$: it is a multiset of sequences of weights, i.e.~a series of $\mathbb{N}\langle\K^{*}\rangle$ mapping each sequence to its multiplicity in the multiset. As usual, we denote multisets via the symbols $\mult{.}$. The disjoint union of two multisets is obtained as the sum of the associated series, it is denoted by $S_1 \cup S_2$. The product of two multisets is obtained as the Cauchy product of the associated series, it is denoted by $S_1 \cdot S_2 = \mult{s_1 s_2 \mid s_1\in S_1, s_2\in S_2}$. This  defines a structure of semiring on multisets where neutral elements are the empty multiset (i.e.~the series mapping all sequences to 0), denoted by $\emptyset$, and the singleton $\mult{\varepsilon}$ that only contains the empty sequence. The constants $\zero$ and $\one$ of the logic represent those constants.  
	
	The semantics of $\WFO$ is defined inductively as follows: 
	\begin{align*}
		\sem{\zero}_{\mathcal{V}}(u,\sigma) & =\emptyset
		\hspace{1cm} \sem{\one}_{\mathcal{V}}(u,\sigma) =\mult{\varepsilon}
		\hspace{1cm} \sem k_{\mathcal{V}}(u,\sigma) =\mult{k}\\
		\sem{\varphi?\Phi_{1}:\Phi_{2}}_{\mathcal{V}}(u,\sigma) & =\begin{cases}
			\sem{\Phi_{1}}_{\mathcal{V}}(u,\sigma) & \text{if }u,\sigma\models\varphi\\
			\sem{\Phi_{2}}_{\mathcal{V}}(u,\sigma) & \text{otherwise}
		\end{cases}\\
		\sem{\Phi_{1}+\Phi_{2}}_{\mathcal{V}}(u,\sigma) & =\sem{\Phi_{1}}_{\mathcal{V}}(u,\sigma)\cup\sem{\Phi_{2}}_{\mathcal{V}}(u,\sigma)\\
		\sem{\Phi_{1}\cdot\Phi_{2}}_{\mathcal{V}}(u,\sigma) & =\sem{\Phi_{1}}_{\mathcal{V}}(u,\sigma)\cdot\sem{\Phi_{2}}_{\mathcal{V}}(u,\sigma)\\
		\sem{\Wexists{x}\Phi}_{\mathcal{V}}(u,\sigma) & =\bigcup_{i\in\{1,2, \ldots,|u|\}}\sem{\Phi}_{\mathcal{V}\cup\{x\}}(u,\sigma[x\mapsto i])\\
		\sem{\WFOrallLR{x}\Phi}_{\mathcal{V}}(u,\sigma) & = \sem{\Phi}_{\mathcal{V}\cup\{x\}}(u,\sigma[x\mapsto 1]) \cdots \sem{\Phi}_{\mathcal{V}\cup\{x\}}(u,\sigma[x\mapsto |u|])
		\\
		\sem{\WFOrallRL{x}\Phi}_{\mathcal{V}}(u,\sigma) & =\sem{\Phi}_{\mathcal{V}\cup\{x\}}(u,\sigma[x\mapsto |u|]) \cdots \sem{\Phi}_{\mathcal{V}\cup\{x\}}(u,\sigma[x\mapsto 1])
	\end{align*}
	
	For sentences (formulas without free variables), we remove the set $\mathcal V$ of variables as well as the valuation $\sigma$ from the notation. 
	Given a series $f\in(\mathbb{N}\langle\mathsf{K}^{*}\rangle)\langle A^{*}\rangle$
	we say that $f$ is \emph{\WFO-definable} if there exists
	a sentence $\Phi_{f}$ such that for all words $u\in A^*$, $f(u)=\sem{\Phi_{f}}(u)$.
	
	We also define the $1$-way fragments $\lrWFO$
	and $\rlWFO$ by discarding binary product
	($\cdot$), as well as $\WFOrallRL{x}$ and $\WFOrallLR{x}$, respectively.
	
	The fragment $\RoneWFO$ of logic studied in \cite{droste2019aperiodic} is obtained by the following grammar:
	\begin{align*}
		\Psi & \Coloneqq k\mid\varphi?\Psi:\Psi & \mathsf{(step\text{-}wFO)}\\
		\Phi & \Coloneqq\zero \mid \varphi?\Phi:\Phi\mid\Phi+\Phi\mid\Wexists{x}\Phi\mid \WFOrallLR{x}\Psi & (\RoneWFO)
	\end{align*}
	where $k\in K$, $\varphi$ is a formula of $\mathsf{FO}$, and $x$ is a first order variable. 
	
	Notice that the abstract semantics of a formula from $\mathsf{step\text{-}wFO}$ maps every word to a singleton multiset. Since $\one$ is removed, as well as the binary product, and $\WFOrallLR{x}$ is restricted to $\mathsf{step\text{-}wFO}$ formulas, it is easy to check inductively that the abstract semantics of a formula from $\RoneWFO$ maps every word $u$ to a multiset of sequences of weights all of the length of $u$. 
	
	To interpret the abstract semantics in terms of a single quantity, we moreover provide an aggregation operator $\aggr\colon \mathbb{N}\langle\mathsf{K}^{*}\rangle \to S$ to a set $S$ of weights. The concrete semantics of a formula $\Phi$ is then obtained by applying $\aggr$ over the multiset obtained via the abstract semantics. The set $S$ can be equipped of various algebraic structures, like semirings or valuation monoids \cite{DroGot11}, as explained in \cite{GasMon14}. In the case of a semiring, we for instance let $\aggr(f)$ be the sum over all sequences $k_1k_2\cdots k_n$ of $f(k_1k_2\cdots k_n) \times k_1 \times k_2 \times \cdots \times k_n$. 
	
	\newcommand{\bfa}{\mathbf a}
	\newcommand{\bfb}{\mathbf b}
	
	\begin{example}\label{ex:logic-mirrorsquared}
		As a first example, consider as a set of weights the languages over the alphabet~$A$. It is naturally equipped with a structure of semiring, where the addition is the union of languages and the multiplication is the concatenation of languages. This semiring is non-commutative which validates our introduction of two product quantification operators, one from left to right and one from right to left. For instance, suppose we want to compute the mapping $f\colon A^* \to 2^{A^*}$ that associates to a word $u$ all the words of the form $\tilde w \tilde w$ with $w$ all factors of $u$ (i.e.~consecutive letters taken inside $u$), where $\tilde w$ denotes the mirror image of the word $w$. For instance, $f(abb) = \{aa, bb, baba, bbbb, bbabba\}$. We can describe this function via a formula of $\WFO$ as follows. We suppose that $A=\{a,b\}$ to simplify, and we let $\K = \{\bfa, \bfb\}$ be the weights that represent the singleton languages $\{a\}$ and $\{b\}$. Then, we describe a formula $\textsf{mirror-factor}(x,y)$ that computes the mirror image of the factor in-between positions pointed by $x$ and $y$: 
		\[\textsf{mirror-factor}(x,y) = \WFOrallRL z \, (x\leq z \land z\leq y)?(P_a(z)?\bfa:\bfb):\one\]
		Then, the mapping $f$ can be described with the formula $\Phi$: 
		\[\Wexists x \Wexists y \, (x \leq y) ? [\textsf{mirror-factor}(x,y) \cdot \textsf{mirror-factor}(x,y)]:\zero \]
		The abstract semantics of the formula associates a multiset of words $\tilde w \tilde w$ with $w$ all factors of $u$. For instance, $\sem{\Phi}(aa) = \mult{\bfa\bfa, \bfa\bfa, \bfa\bfa\bfa\bfa}$. To provide a concrete semantics, we simply consider the aggregation operator that computes the product of sets of weights and removes duplicates in multisets. \qed
	\end{example}
	
	\begin{example}\label{ex:logic-power}
		As a second example, consider the alphabet $A = \{a, b\}$, and the natural semiring $(\mathbb N, +, \times, 0, 1)$, i.e.~the aggregation operator that naturally comes with a semiring. It is a commutative semiring, thus the operator $\WFOrallRL{}$ becomes semantically equivalent (with respect to the concrete semantics, but not to the abstract one) to $\WFOrallLR{}$. Consider the series $f\colon A^* \to \mathbb N$ defined for all words $u\in A^*$ by $f(u) = {|u|_a}^{|u|_b}$, where $|u|_c$ denotes the number of a given letter $c$ in the word $u$. This series can be defined by the following formula, where we intentionally reuse the same variable name twice (but the semantics would be unchanged if the internal variable $x$ was renamed $y$): 
		$\WFOrallLR x \, (P_b(x)? \Wexists x \, (P_a(x)?1:\zero):\one)$. 
		The abstract semantics maps a word with $m$ letters $a$ and $n$ letters $b$ to the multiset containing $m^n$ copies of the sequence $1$. For instance, for the word $abbaa$, the abstract semantics computes $\mult{\varepsilon}\cdot \mult{1, 1, 1}\cdot \mult{1, 1, 1} \cdot \mult{\varepsilon}\cdot \mult{\varepsilon}$, where we have decomposed it with respect to the outermost $\WFOrallLR x$ operator. Once aggregated, all sequences map to $1$, and we thus count $m^n$ as expected.\qed
	\end{example}
	
	\section{Nested Two-Way Weighted Automata}\label{sec:NWA}
	
	Weighted automata are a well-studied model of automata equipped with weights, introduced by Schützenberger~\cite{Sch61}. They have been extended to several weight structures (semirings, valuation monoids), once again with a unified abstract semantics introduced in \cite{GasMon14}. They also have been extended with two-way navigation, and pebbles or nested capabilities, in order to get more power \cite{BolGas14,GasMon18}. In order to simplify our later proofs, we first redefine the semantics of the nested two-way weighted automata with the abstract semantics seen above for the logic.
	
	Since we consider two-way navigation in a word, it is classical to frame the finite word by markers, both on the left and on the right, so that the automaton knows the boundary of the domain. We denote by $\lmark,\rmark$ the left and right markers of the input word, respectively, that are supposed to be symbols not already present in the alphabet $A$ we consider. 
	
	\begin{definition}
		First, by convention, we let $(-1)$-nested two-way weighted automata to be constants of $\K$. 
		Then, for $r\geq 0$, we let $\NWAr r(\K, A)$ (or, $\NWAr r$ if $\K$ and $A$ are clear from the context) be the class of $r$-nested two-way weighted automata over a finite set $\K$ of constants and alphabet $A$, that are all tuples $\A = \langle Q,\trans,I, F\rangle$ where
		\begin{itemize}
			\item $Q$ is a finite set of states;
			\item $\trans$ is the transition relation split into two subsets. 
			\begin{enumerate}
				\item For $a\in A$, there are transitions of the form $(q, a, \B, d, q')\in Q\times A \times \NWAr{(r-1)}(\K, A\times \{0,1\}) \times\{\leftarrow,\rightarrow\}\times Q$, meaning that the automaton is in state $q$, reads the letter $a$, calls the $(r-1)$-nested two-way weighted automaton $\B$ over the same set $\K$ of weights and alphabet $A\times \{0,1\}$ (used to mark the current position), decides to move in the $d$-direction, and changes its state to $q'$.
			
			\item For $a\in A\cup\{\lmark,\rmark\}$, there are some other transitions where the automaton $\B$ is replaced by a weight from $\K$, or by the special constant~$\one$ (that we used in the logic \WFO) to forbid the call of a nested automaton (especially on the  markers): these transitions are thus of the form 
			\begin{align*}
				(q, a, k, d, q') \in &\phantom{{}\cup{}}\big(Q\times A\times (\K\cup \{\one\}) \times\{\leftarrow,\rightarrow\}\times Q\big)\\
				&\cup \big(Q\times \{\lmark\}\times (\K\cup \{\one\}) \times\{\rightarrow\}\times Q\big)\\
				&\cup \big(Q\times \{\rmark\}\times (\K\cup \{\one\}) \times\{\leftarrow\}\times Q\big)
			\end{align*}
			where we have chosen to remove the possibility to move right on a right marker, and left on a left marker (to avoid exiting the possible positions in the input word);
		\end{enumerate}
			\item $I\subseteq Q$ is the set of initial states;
			\item $F\subseteq Q$ is the set of final states. 
		\end{itemize}
	\end{definition}
	
	An automaton $\B$ that appears in the transitions of an automaton $\A$ is called a \textit{child} of~$\A$, and reciprocally $\A$ is a \textit{parent} of $\B$ (notice that an automaton could have several parents, since it can appear in the transitions of several automata). This describes a directed acyclic relationship of dependency between automata: we thus say that an automaton is a \textit{descendant} of another one if they are related by a sequence of parent-child relationship. The unique automaton with no ancestors shall be called the \textit{root}.
	
	In the following a transition of the form $(q, a, x, d, q')$ is said to go from state $q$ to state $q'$, reading letter $a$, having weight $x$, and is called a $d$-transition. 
	
	We now define the abstract semantics of an $\NWAr r(\K, A)$ $\mathcal{A}$, mapping each word $u\in A^*$ to a multiset of sequences of weights $\sem{\A}(u)\in \mathbb{N}\langle\mathsf{K}^{*}\rangle$. Configurations of such an automaton are tuples $(u, i, q)$ where $u= u_1\cdots u_{n}$ is the word in $\{\varepsilon, \lmark\} A^* \{\varepsilon, \rmark\}$ (that could start and end with the markers, or not, in order to be able to define subruns on an unmarked word, that we will use later), $i\in \{1, \ldots, n\}$ is a position in the word,
	and $q\in Q$ is the current state. 
	We call run a sequence $\rho = (u, i_0, q_{0})\xrightarrow{\delta_0, f_0}(u, i_1, q_{1})\xrightarrow{\delta_1, f_1}\cdots\xrightarrow{\delta_m, f_m}(u, i_m, q_{m})$, where  $i_0, \ldots, i_{m-1}\in \{1, \ldots, n\}$, $i_m\in \{0, 1, \ldots, n, n+1\}$ is the final position (that could \textit{exit} the word on left or right)
	$\delta_0, \ldots, \delta_m\in \trans$ and $f_0, \ldots, f_m$ are multisets in  $\mathbb{N}\langle\mathsf{K}^{*}\rangle$ 
	such that for all $j\in\{0, \ldots, m-1\}$:
	\begin{itemize}
		\item $\delta_j$ is a transition from state $q_j$ to state $q_{j+1}$ reading letter $u_{i_j}$;
		\item if $\delta_j$ is a $\rightarrow$-transition then $i_{j+1} = i_j + 1$, otherwise $i_{j+1} = i_j - 1$;
		\item if $u_{i_j} \in A$ and the transition has weight $\B$ that is a $\NWAr{(r-1)}(\K, A\times \{0,1\})$, then $f_j = \sem{\B}(u')$ where $u'$ is the word over alphabet $A\times \{0,1\}$, that will later be denoted by $(u, i_j)$, whose left component is $u$ and whose right component is the constant $0$ except at position $i_j$ where it is $1$;
		\item if the transition has weight $k\in K$, then $f_j = \mult{k}$,
		\item if the transition has weight $\one$, then $f_j = \mult{\varepsilon}$.
	\end{itemize}
	The \emph{initial position} of the run is $i_0$, and its \emph{final position} is $i_m$. The run is \emph{accepting} if $q_0\in I$, $q_m\in F$. Notice that we do not require runs to start on the left marker and stop at the right marker. The weight $\wt(\rho)$ of this run is given as the product of multisets $f_0\cdot f_1 \cdots f_m$.
	
	A run is called \emph{simple} if it never goes twice through the same configuration. Not all runs are simple, but we restrict ourselves to using only those in the semantics: otherwise, an infinite number of runs should be summed, which would produce an infinite multiset (and then the aggregator function should be extended to add this possible behaviour). This restriction was also considered in \cite{BolGas14,GasMon18}.
	
	We then let $\sem{\A}(u)$ be the union (as multiset) of the weights of accepting simple runs (whatever their initial and final positions). As for the logics above, we may then use an aggregation operator to obtain a \emph{concrete semantics} $\csem{\A}$ mapping each word $u$ to a weight structure $S$.
	
	Given a series $f\in(\mathbb{N}\langle\mathsf{K}^{*}\rangle)\langle A^{*}\rangle$
	we say that $f$ is \emph{$\NWA$-definable} if there exists $r\geq 0$ and an $\NWAr r(\K, A)$ $\A$
	such that for all words $u\in A^*$, $f(u) = \sem{\A}(u)$.
	
	\begin{figure}[tbp]
		\centering
		\scalebox{.8}{\begin{tikzpicture}[node distance=3.5cm]
				\node[state,initial](0) {} node[left of=0,node distance=1cm](){$\A$};
				\node[state,right of=0,accepting right,accepting by arrow] (1) {};
				\draw[->] 
				(0) edge node[above]{$A,\A_x,{\rightarrow}$} (1)
				;
				
				\begin{scope}[xshift=6.5cm]
					\node[state,initial](0) {} node[left of=0,node distance=1cm](){$\A_x$};
					\node[state,right of=0] (1) {};
					\node[state,right of=1,accepting right,accepting by arrow] (2) {};
					\node[state,above of=1,node distance=2cm] (1a) {};
					\node[state,below of=1,node distance=9mm] (1b) {};
					
					\draw[->] 
					(0) edge node[above]{$(A,1),\one,{\rightarrow}$} (1)
					(1) edge[loop above] node[above]{$(A,0),\one,{\rightarrow}$} (1)
					(1) edge node[above]{$(A,0),\A_{x,y},{\rightarrow}$} (2)
					(0) edge[bend left=10] node[above left] {$(a,1),\{a\},{\rightarrow}$} (1a)
					(1a) edge[bend left=10] node[above right,xshift=-5mm,yshift=2mm] {$\begin{array}{c}
					\rmark, \{a\}, {\leftarrow} \\ (A, 0), \{a\}, {\rightarrow}\end{array}$} (2)
					(0) edge[bend right=10] node[below,xshift=-5mm] {$(b,1),\{b\},{\rightarrow}$} (1b)
					(1b) edge[bend right=10] node[below,xshift=5mm,yshift=-2mm] {$\begin{array}{c}
					\rmark, \{a\}, {\leftarrow} \\ (A,0), \{b\}, {\rightarrow}\end{array}$} (2)
					; 
				\end{scope}
				
				\begin{scope}[yshift=-3cm,xshift=2cm,node distance=3.5cm]
					\node[state,initial] (1) {} node[left of=1,yshift=-7mm,node distance=1cm](){$\A_{x,y}$};
					\node[state,right of=1] (2) {};
					\node[state,right of=2,yshift=-7mm] (3) {};
					\node[state,left of=3,yshift=-7mm] (4) {};
					\node[state,left of=4,accepting left,accepting by arrow] (5) {};
					
					\draw[->] 
					(1) edge node[above]{$(a,0,1),\{a\},{\leftarrow}$} node[below]{$(b,0,1),\{b\},{\leftarrow}$} (2)
					(2) edge[loop above] node[above]{$\begin{array}{c}(a,0,0),\{a\},{\leftarrow}\\(b,0,0),\{b\},{\leftarrow}\end{array}$} (2)
					(2) edge node[above,xshift=6mm]{$\begin{array}{c}(a,1,0),\{a\},{\rightarrow}\\(b,1,0),\{b\},{\rightarrow}\end{array}$} (3)
					(3) edge[loop right] node[right]{$(A,0 , 0),\one,{\rightarrow}$} (3)
					(3) edge node[below,xshift=6mm]{$\begin{array}{c}(a,0,1),\{a\},{\leftarrow}\\(b,0,1),\{b\},{\leftarrow}\end{array}$} (4)
					(4) edge[loop below] node[below]{$\begin{array}{c}(a,0,0),\{a\},{\leftarrow}\\(b,0,0),\{b\},{\leftarrow}\end{array}$} (4)
					(4) edge node[above]{$(a,1,0),\{a\},{\leftarrow}$} node[below]{$(b,1,0),\{b\},{\leftarrow}$} (5)
					; 
				\end{scope}
		\end{tikzpicture}}
		\caption{An $\NWAr 2$ that recognises the series described in Example~\ref{ex:logic-mirrorsquared}. The letter $A$ is used in transitions to denote the presence of all possible letters from $A$. Notice that the runs of all the automata may start and stop at any position of the word, as described in the semantics.}
		\label{fig:auto-mirrorsquared}
	\end{figure}
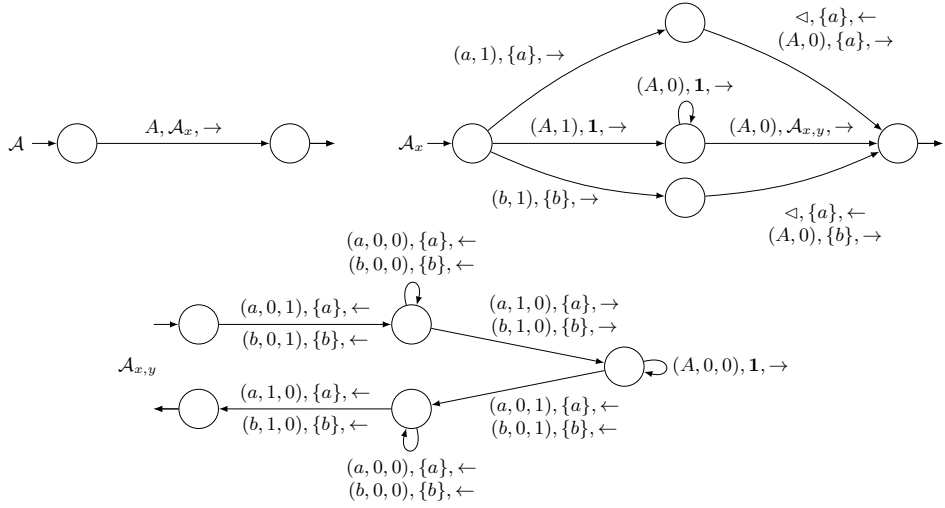
	
	\begin{example}
		We describe in Figure~\ref{fig:auto-mirrorsquared} a $\NWAr 2$  $\A$ that recognises the series described in Example~\ref{ex:logic-mirrorsquared}. Two levels of nesting are used to mark non-deterministically the positions $x$ and $y$, such that $x\leq y$ (or only one of them if $x=y$). Then, the last level of nesting is used to compute the value of the formula $\textsf{mirror-factor}(x,y) \cdot \textsf{mirror-factor}(x,y)$ by two passes from right to left. \qed
	\end{example}
	
	A run over a word $u$ is called \textit{left-to-right} (resp.~\textit{left-to-left}, \textit{right-to-right}, \textit{right-to-left}) if its initial position is $1$ (resp.~$1$, $|u|$, $|u|$) and its final position is $|u|+1$ (resp.~$0$, $|u|+1$, $0$). Intuitively, we thus use this terminology to detect if a run starts on the first or last position of the word, and if it exits the word either on the left or on the right.
	
	\noindent \textbf{Navigational restrictions.}
	An $r\text{-}\lrNWA$ (respectively, $r\text{-}\rlNWA$) is an $\NWAr r$ where all transitions appearing in the automaton or its descendants are $\rightarrow$-transitions (resp.~$\leftarrow$-transitions). Those models are called \emph{one-way} in the following, since the head movement is fixed during the whole run. 
	
	An $r$-nested \textit{sweeping} weighted automaton ($r\text{-}\swNWA$) is an $\NWAr r$ where changes of directions are only allowed (in this automaton or its descendants) at  markers. More formally, states of the automaton and each of its descendants are separated in two sets $Q^{\rightarrow}$ and $Q^{\leftarrow}$ such that for all transitions $(q, a, \B, d, q')$ or $(q, a, k, d, q')$, 
	\begin{itemize}
		\item if $q, q'\in Q^{\rightarrow}$, then $d={\rightarrow}$; 
		\item if $q, q'\in Q^{\leftarrow}$, then $d={\leftarrow}$; 
		\item if $q\in Q^{\rightarrow}$ and $q'\in Q^{\leftarrow}$, then $d={\leftarrow}$ and $a = \rmark$;
		\item if $q\in Q^{\leftarrow}$ and $q'\in Q^{\rightarrow}$, then $d={\rightarrow}$ and $a = \lmark$. 
	\end{itemize} 
	
	\noindent \textbf{Ambiguity.}
	An $\NWAr{r}$ $\A$
	is called \textit{polynomially ambiguous} if there is a polynomial $p$ such
	that over every word $u$ on its alphabet the number of accepting runs of $\A$, as well as the number of accepting runs of any of its descendants, is at most $p(|u|)$. If the polynomial $p$ is linear, $\A$ is said to be \emph{linearly ambiguous}. If the polynomial $p$ is the constant $1$, $\A$ is said to be \emph{unambiguous}. Notice that the condition deals with all runs, and not only the simple ones.
	
	Polynomial ambiguity (indeed even \emph{finite} ambiguity, where the number of accepting runs must be finite for all words) implies that all accepting runs are simple: otherwise, there would be an infinite number of accepting runs, by allowing the loops to happen as many times as possible. 

	The $\NWAr 2$ of Figure~\ref{fig:auto-mirrorsquared} is linearly ambiguous since the toplevel automaton~$\A$ has only to choose the position where to start the run, the automaton $\A_x$ has then only to choose the position where to call the next automaton, and the automaton $\A_{x,y}$ is unambiguous.
	
	\noindent \textbf{Aperiodicity.}
	In order to define a notion of aperiodicity for $\NWA$, we need to enhance the usual notion of aperiodicity for automata to incorporate weights, two-way navigations, and nesting. As in \cite{droste2019aperiodic}, we simply do not care about weights and thus require that the unweighted version of the automata are aperiodic. For two-way navigations, we rely on existing extensions of the notion of \emph{transition monoid} for two-way automata and transducers~\cite{birget1989concatenation,birget1990two,carton2021aperiodic}. Finally, for nesting, we simply require that each automaton appearing in an \NWA is aperiodic. 
	
	More formally, given a $\NWA$ $\mathcal{A}$ over the alphabet $A$, its transition monoid is the quotient of the free monoid $A^*$ by a congruence relation capturing the equivalence of two \emph{behaviours} of the automaton. As for runs, we distinguish four types of behaviours: left-to-left, left-to-right, right-to-left and right-to-right. The left-to-left behaviour $\bh{\mathcal{A}}{ll}(w)$ of $w\in \{\varepsilon,\lmark\}A^*\{\varepsilon,\rmark\}$ in $\mathcal{A}$ is the set of pairs of states $(p, q)$ such that there exists a left-to-left run over $w$ from state $p$ to state~$q$ (notice that we do not care if the descendant automata that are called along this run are indeed "accepting" the word). The other behaviours can be defined analogously.
	
	\begin{definition}
		Let $\mathcal{A} = \tup{Q,\trans,I,F}$ be a $\NWA(\K,A)$. The transition monoid of~$\mathcal{A}$ is $A^*\setminus\sim_{\mathcal{A}}$ where $\sim_{\mathcal{A}}$ is the conjunction of the following congruence relations, defined for $w, w'\in A^*$ by:
		\begin{itemize}
			\item $w\sim^{\mathcal{A}}_{ll}w'$ iff $\bh{\mathcal{A}}{ll}(w)=\bh{\mathcal{A}}{ll}(w')$
			\item $w\sim^{\mathcal{A}}_{lr}w'$ iff $\bh{\mathcal{A}}{lr}(w)=\bh{\mathcal{A}}{lr}(w')$
			\item $w\sim^{\mathcal{A}}_{rl}w'$ iff $\bh{\mathcal{A}}{rl}(w)=\bh{\mathcal{A}}{rl}(w')$
			\item $w\sim^{\mathcal{A}}_{rr}w'$ iff $\bh{\mathcal{A}}{rr}(w)=\bh{\mathcal{A}}{rr}(w')$
		\end{itemize}
	\end{definition}
	
	Notice that in the previous definition, we only focus on words not containing markers. This is because we only use this monoid in order to define aperiodicity of the automata where we focus on powers of elements of the monoid, which correspond to runs on iterates of a word in which it makes no sense to duplicate some markers.  
	
	An $\NWAr{r}$ is \emph{aperiodic} if its transition monoid, as well as the ones of all its descendants, are aperiodic (i.e.~for all elements $x$ of the monoid, there is a natural number $n$ such that $x^n = x^{n+1}$).
	
	Given an $\NWA$ $\A$, its \emph{left-to-right} (resp.~\emph{right-to-left}) projection is the $\NWA$~$\overrightarrow{\A}$ (resp. $\overleftarrow \A$) obtained by only keeping $\rightarrow$-transitions (resp.~$\leftarrow$-transitions) in the root automaton. 
	Interestingly, when starting from sweeping automata, aperiodicity is preserved when taking such projections.
	
	\begin{lemma}
		\label{lem:aperiodicity-of-projections} If a $\swnwA$ $\A$ is aperiodic
		then $\overrightarrow{\A}$ and $\overleftarrow{\A}$ are aperiodic.
	\end{lemma}
	\begin{proof}
		Let $\A=\tup{Q,\trans,I,F}$ be a $\swnwA(K, A)$. We prove the result for $\overrightarrow{\A}$. The proof for $\overleftarrow{\A}$ follows analogously.
		
		Consider the word $u\in A^*$. Then there exists a natural number $k$ such that $\bh{\A}{e}(u^k)=\bh{\A}{e}(u^{k+1})$ for $e\in \{ll,lr,rl,rr\}$. 
		If $u=\epsilon$, then $\bh{\overrightarrow{\A}}{e}(u)=\{(p,p)\mid p\in Q\}=\bh{\overrightarrow{\A}}{e}(u^2)$ for $e\in \{ll,lr,rl,rr\}$. It remains to prove the lemma when $u$ is not empty. Immediately, we have $\bh{\overrightarrow{\A}}{ll}(u)=\bh{\overrightarrow{\A}}{ll}(u^2)=\bh{\overrightarrow{\A}}{rl}(u)=\bh{\overrightarrow{\A}}{rl}(u^2)=\emptyset$. Since no run of the sweeping automaton $\A$ over $u$ can change the direction of its head movement over $w$ that does not contain end markers, we have $\bh{\overrightarrow{\A}}{lr}(u^{k})=\bh{\A}{lr}(u^k)=\bh{\A}{lr}(u^{k+1})=\bh{\overrightarrow{\A}}{lr}(u^{k+1})$. Finally, $\bh{\overrightarrow{\A}}{rr}(u) = \{(p,q)\mid (p,u_{|u|},x,{\rightarrow},q)\in\trans\} = \bh{\overrightarrow{\A}}{rr}(u^2)$.
		
		For every word $u\in A^*$, we thus have $\bh{\overrightarrow{\A}}{e}(u^{k'})=\bh{\overrightarrow{\A}}{e}(u^{k'+1})$ for $e\in \{ll,lr,rl,rr\}$, where $k'=\max(1,k)$. Hence, $\overrightarrow{\A}$ is aperiodic.
		\qed\end{proof}
	
	\noindent \textbf{Results.}
	The goal is to find an adequate characterisation of the automata models that recognise exactly the series \twoWFO-definable or the fragments introduced before. Droste and Gastin~\cite{droste2019aperiodic} paved the way of this study by characterising \RoneWFO with a fragment of the classical one-way weighted automata, that are $0$-$\lrNWA$ where the semantics is computed by only considering accepting runs that start on the first letter of the word, and end on the last letter of the word: this is because of the specific type of multiset produced by formulas of \RoneWFO, all elements being of the same length (the length of the input word). In particular, markers are useless in this context.
	
	\begin{theorem}[\cite{droste2019aperiodic}]
		For all series $f\in(\mathbb{N}\langle\mathsf{K}^{*}\rangle)\langle A^{*}\rangle$, the following conditions are equivalent: 
		\begin{enumerate}
			\item $f$ is definable by a polynomially ambiguous aperiodic $0$-$\lrNWA$
			\item $f$ is \RoneWFO-definable.
		\end{enumerate}
	\end{theorem}
	
	Here, and in the following, classes $\mathcal C_1$ and $\mathcal C_2$ of models are said to be equivalent if for all models $\mathcal M_1\in \mathcal C_1$ and $\mathcal M_2\in \mathcal C_2$ working on the same alphabet and the same sets of weights, and for all words $u$, the abstract semantics $\sem{\mathcal M_1}(u)$ is equal to the abstract semantics $\sem{\mathcal M_2}(u)$. This then implies that, for every aggregation function, the concrete semantics are also the same. 
	
	The proof of this theorem is constructive in both directions, and we will revisit it in the next sections, providing generalisations of it in order to get our main contribution: 
	
	\begin{theorem}\label{thm:contribution-2way}
		For all series $f\in(\mathbb{N}\langle\mathsf{K}^{*}\rangle)\langle A^{*}\rangle$, the following conditions are equivalent: 
		\begin{enumerate}
			\item\label{item:linear-2way} $f$ is definable by a linearly ambiguous aperiodic \NWA;
			\item\label{item:2way} $f$ is definable by a polynomially ambiguous aperiodic \NWA;
			\item\label{item:logic} $f$ is \twoWFO-definable;
			\item\label{item:linear-sweeping} $f$ is definable by a linearly ambiguous aperiodic \swNWA
			\item\label{item:sweeping} $f$ is definable by a polynomially ambiguous aperiodic \swNWA.
		\end{enumerate}
	\end{theorem}
	
	The rest of the article is devoted to a sketch of the proofs of Theorem~\ref{thm:contribution-2way}. We provide in Section~\ref{sec:logic2aut} the sketch of proof of $\ref{item:logic}\Rightarrow\ref{item:linear-sweeping}$, in Section~\ref{sec:sweeping2logic} the sketch of proof of $\ref{item:sweeping}\Rightarrow\ref{item:logic}$, and in Section~\ref{sec:2way2sweeping} the sketch of proof of $\ref{item:2way}\Rightarrow\ref{item:sweeping}$. We can then conclude by the trivial implications $\ref{item:linear-sweeping}\Rightarrow \ref{item:linear-2way}\Rightarrow\ref{item:2way}$. 
	
	As a side result, we also obtain a characterisation for one-way models:  
	\begin{theorem}\label{thm:contribution-1way}
		For all series $f\in(\mathbb{N}\langle\mathsf{K}^{*}\rangle)\langle A^{*}\rangle$, the following conditions are equivalent: 
		\begin{enumerate}
			\item\label{item:linear-1way} $f$ is definable by a linearly ambiguous aperiodic \lrNWA;
			\item\label{item:1way} $f$ is definable by a polynomially ambiguous aperiodic \lrNWA;
			\item\label{item:1way-logic} $f$ is \lrWFO-definable.
		\end{enumerate}
	\end{theorem}

	These theorems complete the picture initiated in \cite[Theorem~5.11]{BolGas14} where it is shown that, in commutative semirings, \NWA (called \emph{pebble two-way weighted automata}, with a more operational view of dropping/lifting pebbles, but the expressive power is identical) are equivalent to an extension of the logic \twoWFO with a bounded weighted transitive closure operator. It is also noted that, even in non commutative semirings, the whole logic \lrWFO with the bounded transitive closure operator can be translated into equivalent \NWA.
	
	\section{From the Logic to Automata}\label{sec:logic2aut}
	
	In this section, we prove the implication $\ref{item:logic}\Rightarrow\ref{item:linear-sweeping}$ of Theorem~\ref{thm:contribution-2way}. 
	This is obtained by a generalisation of the proof given by Droste and Gastin in \cite[Theorem~16]{droste2019aperiodic}, where they only deal with restricted one-way logic and non-nested one-way automata. The proof is indeed simpler since we can rely on the use of nesting, contrary to them where they need a careful construction for formulas $\WFOrallLR{x}\Psi$
	of $\RoneWFO$.
	
	The construction is performed by induction on the formula of \WFO, making use of nesting in automata, as originally demonstrated in \cite[Proposition~5.13]{BolGas14} to transform every formula of a logic containing $\WFO$ (as well as a bounded weighted transitive closure operator) into $\NWA$.
	
	As known since \cite{Sch65,McNPap71}, from every formula $\varphi$ of \FO, we can obtain an equivalent classical deterministic finite state automaton that is aperiodic, starts on the marker $\lmark$ and ends on the marker $\rmark$. By putting on every transition the weight $\one$, this results in a $\lrNWAr 1$ $\A_\varphi$ that is unambiguous and aperiodic, whose abstract semantics is equal to the formula $\varphi?\one:\zero$.
	
	Consider then a formula $\Phi=\varphi?\Phi_{1}:\Phi_{2}$, where, by induction, we already have two $\NWAr{r}$ $\A_1$ and $\A_2$ for $\Phi_{1}$ and $\Phi_{2}$ (without loss of generality we adapt the maximal level $r$ of nesting by adding useless levels). We can use the $\lrNWAr 1$~$\A_\varphi$ in order to produce an $\NWAr{r}$ equivalent to $\Phi$: the first level consists in $\A_\varphi$, and once it unambiguously reach the marker $\rmark$, we continue the run by going back to the left marker, and continue either to $\A_1$ or to $\A_2$, whether the formula~$\varphi$ was concluded to be satisfied or not, respectively.\footnote{In the proof of Theorem~\ref{thm:contribution-1way}, we replace this construction by the use of nesting that allows one to restart from the first position of the word in order to compute the behaviour of either $\A_1$ or $\A_2$.}
	
	The sum and product of two formulas can be computed by taking the disjoint union of two automata, or by starting the computation of the second after the computation of the first one (either by going back to the beginning of the word, or using a level of nesting). 
	
	For the quantification operators, we use one more level of nesting. Suppose that we have an $\NWAr{r}(\K, A\times \{0, 1\})$ $\A$ equivalent to a formula $\Phi$ with a free variable $x$. Then, the formula $\Wexists x \, \Phi$ can be defined by the following $\NWAr{(r+1)}(\K, A)$, making use of the fact that we can non-deterministically start and end a run wherever we want: the automaton thus has a single transition that calls $\A$. 
	For the $\WFOrallLR{x}$ operator, the toplevel automaton scans the whole word from left to right, and calls $\A$ on each position (that is not a marker). For the $\WFOrallRL{x}$ operator, we do the same but starting from the right marker and scanning the whole word from right to left.
	In both the cases, the root of the resulting automaton is aperiodic. 
	
	To conclude that the constructed $\NWA$ is linearly ambiguous and aperiodic, we make use of the fact that linearly ambiguous aperiodic automata are closed under disjoint union, nesting and concatenation with unambiguous (even finitely ambiguous) automata. It is indeed true for the case of disjoint unions, the individual automata still preserve the aperiodicity in their simulations and any run in the new automaton must be restricted to one of the automata. In the case of nesting, every transition of the \emph{soon-to-be-child} automaton is replaced by all its extensions with respect to the input letter. Since the transitions are now oblivious to the marking of an input, the aperiodicity of the new child automaton is once again ensured under the extended alphabet. To understand the closure of linear ambiguity of automata under concatenation with unambiguous automata, one must just observe that the concatenation of automata essentially multiples the ambiguities of the factor automata, since every run in the concatenation is the sequence of a run in the first factor and one in the second.

	\section{From Nested Sweeping Weighted Automata to the Logic}\label{sec:sweeping2logic}
	
	This section aims at proving the implication $\ref{item:sweeping}\Rightarrow\ref{item:logic}$ of Theorem~\ref{thm:contribution-2way}. 
	We shall first
	prove it in the $1$-way case.
	
	\begin{lemma}\label{lem:aut-to-logic-1way}
		For all polynomially ambiguous aperiodic $\lrNWAr r$ (resp.~$\rlNWAr r$), 
		there exists an equivalent formula of \lrWFO (resp.~\rlWFO).
	\end{lemma}
	
	\begin{proof}
		Once again, we only deal with the left-to-right result, the other one being obtained symmetrically.
		Let $\A_{p,q}$ denote the $\lrNWAr{r}$ obtained from $\A$ where the initial and final states are $p$ and $q$ respectively. We prove by induction on $r$, that for all $\lrNWAr r$ $\A$ and each pair of states $p$
		and $q$, we can construct a $\lrWFO$ sentence
		$\Phi_{p,q}$ such that $\sem{\A_{p,q}}=\sem{\Phi_{p,q}}$. We then conclude by considering all initial states $p$ and final states $q$.
		
		If $r=0$, the result follows, after trimming the root of $\A$ so that all states can be reached from an initial state and reach a final state (no matter if the descendant automata called on the transition indeed accept the word), directly from the construction of Droste and Gastin~\cite[Proposition 9 and Theorem 10]{droste2019aperiodic}. Note that trimming the automaton does not alter its semantics. The main difference in our case is the fact that our automata can non-deterministically start and end in the middle of the word. We may however start by modifying them to force them to start on the left marker and end on the right marker: it suffices to add self-loop transitions at the beginning and the end of weight $\one$ (so that these additional transitions do not modify the abstract semantics).
		
		We now suppose that $r>0$, and assume that the result holds for $r-1$. 
		Consider an $\lrNWAr{r}(\K,A)$ $\A$ that we suppose trimmed. As in the previous case, we can produce a formula $\Phi$ for $\A$, abstracting away for now the weight $k_\B$ on the transitions that stands for a $\lrNWAr{(r-1)}(\K,A\times \{0,1\})$ $\B$. 
		
		We use the induction hypothesis to produce a formula $\Phi_\B$ of $\WFO$ for every $\lrNWAr{(r-1)}(\K,A\times \{0,1\})$ $\B$ that appears in the transitions of $\A$. We modify this formula so that we incorporate a fresh first order variable $x$ standing for the position on which $\B$ is called. Then, we replace every subformula $P_{(a,i)}(y)$ with $(a,i)\in A\times \{0,1\}$ by $P_a(y)\land y=x$ if $i=1$, $P_a(y)\land y\neq x$ if $i=0$. 
		
		In the formula $\Phi$ produced by Droste and Gastin, each weight $k_\B$ appears in a subformula with a distinguished first order variable $x$ encoding the position of the letter read by the transition that should compute the weight $k_\B$. Thus, we simply replace every such weight $k_\B$ by the modified formula $\Phi_\B$ above.
		\qed\end{proof}
	
	We then turn to the case of sweeping automata. 
	
	\begin{lemma}\label{lem:base-case-aut-to-logic-full}
		For every polynomially ambiguous aperiodic \swNWA, there exists an equivalent formula of \WFO. 
	\end{lemma}
	
	\begin{proof}
		The proof also goes by induction on the level of nesting, and follows the same construction as the previous lemma. The only novelty is the treatment of change of directions in the runs. We thus only consider the case of $\swNWAr 0$ below. 
		
		For a $\swNWAr 0$ $\A = \tup{Q,\trans,I,F}$, 
		we show that for each pair of states $p$ and $q$, we can
		construct a formula of \WFO $\Phi_{p,q}$ equivalent to $\A_{p,q}$. As before, without loss of generality, we can suppose that every accepting run starts on the left marker, and stops on the right marker.
		
		Given a word $w=w_{1}\cdots w_{m}$, every run from $p$ to $q$ on
		$\word{w}$ can then be decomposed as 
		\begin{multline*}(p, \lmark, k_0, {\rightarrow}, p_{0})\rho_{1} (p_{1}, \rmark, k_2, {\leftarrow}, p_{2}) \rho_{2} (p_{3}, \lmark, k_4, {\rightarrow}, p_{4})\cdots \\ 
			(p_{2n-1}, \lmark, k_{2n}, {\rightarrow}, p_{2n})\rho_{2n+1}
			(p_{2n+1}, \rmark, k_{2n+2}, {\leftarrow}, q)
		\end{multline*}
		where $\rho_{2i+1}$ only contains $\rightarrow$-transitions (for $i\in \{0, \ldots, n\}$), and $\rho_{2i}$ only $\leftarrow$-transitions (for $i\in \{1, \ldots, n\}$). 
		Since we assume polynomial
		ambiguity of $\A$, we must have $n\leq|Q|$. Otherwise, there exists a position which is visited twice in the same state, thus allowing infinitely many runs over the input word by pumping this looping fragment of the run.
		We then immediately obtain that, for every word $w$, the multiset $\sem{\mathcal{A}_{p,q}}(w)$ can be decomposed as
		\[\sum_{\substack{n\leq|Q|\\(p, \lmark, k_0, {\rightarrow}, p_{0}), \ldots,\\ (p_{2n+1}, \rmark, k_{2n+2}, {\leftarrow}, q)\in \trans}}
		\!\!\!\!\!\!\!\!\!\!\!\!\!\!\!
		\mult{k_0}\sem{\overrightarrow{\mathcal{A}}_{p_0,p_{1}}}(w)\mult{k_2}\sem{\overleftarrow{\mathcal{A}}_{p_{2},p_3}}(w)
		\cdots \sem{\overrightarrow{\mathcal{A}}_{p_{2n},p_{2n+1}}}(w)\mult{k_{2n+2}}
		\]
		
		It remains to show that the above decomposition can be translated
		into an equivalent $\WFO$ sentence. Since trimming preserves aperiodicity, using Lemma~\ref{lem:aperiodicity-of-projections}, we know that
		for every $p,q\in Q$, both $\overrightarrow{\A}_{p,q}$ and $\overleftarrow{\A}_{p,q}$
		are aperiodic. By Lemma~\ref{lem:aut-to-logic-1way}, we can thus construct equivalent $\WFO$ sentences $\overleftarrow{\Phi}_{p,q}$
		and $\overrightarrow{\Phi}_{p,q}$, respectively.
		We now define, 
		\[
		\Phi_{p,q} =\!\!\!\!\!
		\sum_{\substack{n\leq|Q|\\(p, \lmark, k_0, {\rightarrow}, p_{0}), \ldots,\\ (p_{2n+1}, \rmark, k_{2n+2}, {\leftarrow}, q)\in \trans}}
		\!\!\!\!\!\!\!\!\
		k_0\cdot \overrightarrow{\Phi}_{p_0,p_{1}}\cdot
		k_2
		\cdot\overleftarrow{\Phi}_{p_{2},p_{3}}\cdots 
		\cdot\overrightarrow{\Phi}_{p_{2n},p_{2n+1}}\cdot k_{2n+2}
		\]
		It can be proved that $\sem{\mathcal{A}_{p,q}}=\sem{\Phi_{p,q}}$. 
		Finally, we set $\Phi=\sum_{p\in I,q\in F}\Phi_{p,q}$ and we can check that this formula is equivalent to $\A$. 
		\qed
	\end{proof}
	
	\section{From Nested Two-Way Weighted Automata to Nested Sweeping Weighted Automata}\label{sec:2way2sweeping}
	
	In this section we finally provide a sketch of the proof of $\ref{item:2way}\Rightarrow\ref{item:sweeping}$ in Theorem~\ref{thm:contribution-2way}. 
	This is the most novel and challenging part of the proof. In particular, notice that such an implication requires the use of nesting: the following example of $\NWAr 0$ does not have an equivalent $\swNWAr 0$, even under the restriction
	of polynomial ambiguity and aperiodicity. 
	
	\begin{figure}[tbp]
		\centering
		\begin{tikzpicture}[node distance=3cm]
			\node[state] (p) {$p$};
			\node[state, initial] (init) [left of=p,yshift=.8cm] {$\iota$};
			\node[state, accepting by arrow, accepting left] (fin) [left of=p,,yshift=-.8cm] {$\kappa$};
			\node[state] (q) [right of=p] {$q$};
			\node[state] (r) [right of=q] {$r$};
			
			\draw[->] (init) edge node[above, xshift=3mm,pos=.3] {$\lmark, \one, \rightarrow$} (p) 
			(p) edge node[below, xshift=3mm,pos=.7] {$\rmark,\one,\leftarrow$} (fin)
			(p) edge [loop above] node [above] {$a, f, \rightarrow$} ();
			\draw[->] (p) edge [] node [align=center] {$b, \one, \leftarrow$} (q);
			\draw[->] (q) edge [loop above] node [align=center, above] {$a, g, \leftarrow$} ();
			\draw[->] (r) edge [bend left=20] node [align=center, below] {$b, \one, \rightarrow$} (p);
			\draw[->] (r) edge [loop right] node [right] {$a, \one, \rightarrow$} ();
			\draw[->] (q) edge node [align=center, above] {$\begin{array}{c}\lmark, \one, \rightarrow\\ b, \one, \rightarrow\end{array}$} (r);
		\end{tikzpicture}
		\caption{\label{fig:counter-eg-same-layer} A $\NWAr{0}$ $\A_{ex}$.}
	\end{figure}
	\begin{example}\label{ex:running}
		Consider the $\NWAr 0(\K, A)$ $\A_{ex}$ depicted in
		Figure~\ref{fig:counter-eg-same-layer}, over the alphabet $A=\{a,b\}$
		and with weights $\K = \{f, g\}$. Its semantics maps every word of $A^*$ of the form
		$u=a^{m_{1}}b\cdots a^{m_{n}}b$ to the multiset 
		$\mult{f^{m_{1}}g^{m_{1}}\cdots f^{m_{n}}g^{m_{n}}}$, and every word of the form
		$u=a^{m_{1}}b\cdots a^{m_{n}}$ to the multiset 
		$\mult{f^{m_{1}}g^{m_{1}}\cdots f^{m_{n}}}$.
		The automaton $\A_{ex}$ is unambiguous (even deterministic). By a computation of its transition
		monoid, it can also be shown to be aperiodic. 
		We can prove 
		that it has no equivalent $\swNWAr{0}$, since it is crucial that the automaton switches direction several times in the middle of the word.\qed
	\end{example}
	
	Consider now a $\NWAr 0$ $\A$ (we will explain at the very end how to do this for an $\NWAr r$). We build an $\swNWA$ $\overline\A$ equivalent to it, that will moreover be aperiodic and polynomially ambiguous if $\A$ is.
	
	To understand our construction of $\overline \A$, consider an accepting simple run of $\A$ over a
	word $\lmark u\rmark$ startingfrom the left marker and ending at the right marker\footnote{As seen in Lemma \ref{lem:aut-to-logic-1way} in a special case, we can always massage the automata to ensure that all accepting runs start from the left marker and end at the right marker, while preserving aperiodicity.}. In order to get closer to a sweeping automaton, we first split the run into subruns that go from the left marker to the right marker (possibly hitting in the mean time the left marker), and then to the left marker again (possibly hitting in the mean time the right marker), and so on. We get at most $|Q|$ subruns by doing so, since the run is simple (and thus cannot visit more than $|Q|$ times each marker).
	
	For each subrun, we further decompose them as follows. We only present here the decomposition for the left-to-right case, the other one being symmetrical. 
	
	For a left-to-right run over the word $w=w_{1}\cdots w_{n}$ (with $w_1$ possibly being equal to $\lmark$, but $w_n\neq \rmark$), we decompose it into the interleaving of subruns with only $\rightarrow$-transitions, ending in an increasing sequence of positions $(i_j)_{1\leq j\leq m}$, and some right-to-right subruns on the prefix words $w_1 \cdots w_{i_j}$. 
	Formally, every left-to-right run can be written as
	$\rho_1 \lambda_1 \rho_2 \lambda_2 \cdots \lambda_{m-1} \rho_m$
	where we have a sequence of positions $0=i_{0}<i_{1}<\cdots<i_{m-1}<i_{m}=n+1$ such that
	\begin{itemize}
		\item for all $j\in \{1, \ldots, m\}$, $\rho_j$
		is a run over $w_{i_{j-1}+1}\cdots w_{i_{j}-1}$ with only $\rightarrow$-transitions: notice that this run can be empty if $i_{j}=i_{j-1} + 1$;
		\item for all $j\in \{1, \ldots, m-1\}$, $\lambda_j$ is a right-to-right run over 
		$w_{1}\cdots w_{i_j}$ that starts with a $\leftarrow$-transition.
	\end{itemize}
	
	\begin{figure}[tbp]
		\centering
		\includegraphics[scale=.25,page=1]{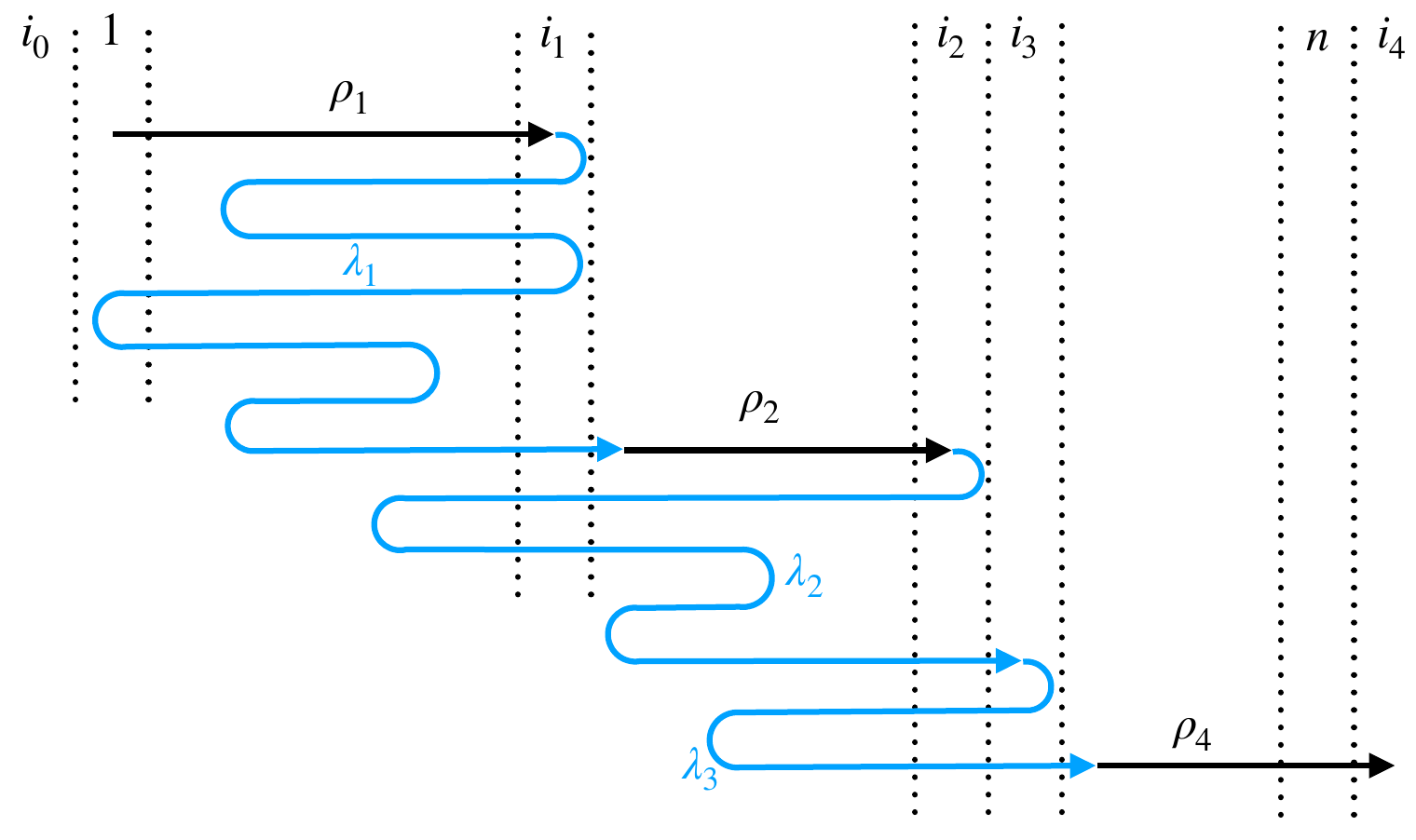}\qquad
		\raisebox{6.5mm}{\includegraphics[scale=.25,page=2]{runs}}
		\caption{On the left, the decomposition of a left-to-right run as a sequence $\rho_1\lambda_1\rho_2\lambda_2\rho_3\lambda_3\rho_4$ with $\rho_3$ being empty. On the right, the decomposition of a right-to-right run as a sequence $\rho_1\lambda_1\rho_2\lambda_2\rho_3\tau$.\label{fig:abstract-decomposition}}
	\end{figure}
	
	We exemplify the decomposition on the left of Figure~\ref{fig:abstract-decomposition}. We thus build an \NWA $\overline\A$ whose root automaton is a sweeping automaton that emulates the black $\rho$-parts, interleaved with some new $\rightarrow$-transitions from state $p$ (in a position $i_j$) to state $q$ (in the corresponding position $i_j+1$): the weight of this transition is a (non-sweeping) $\NWA(\K, A\times \{0,1\})$ $\A_{(p,q)}$ that is in charge of emulating the blue subrun $\lambda_j$ from state $p$ to state $q$ of $\A$, keeping marked the position $i_j$ in the second component of the alphabet $A\times \{0,1\}$. 
	
	We treat the various new automata $\A_{(p,q)}$ recursively to transform them to sweeping automata too. We thus similarly decompose the $\lambda$-subruns as before, by adapting the previous decomposition working only for left-to-right runs. 
	In the decomposition of a right-to-right run over the word $w=w_{1}\cdots w_{n}$ (with $w_1$ possibly being equal to $\lmark$, but $w_n\neq \rmark$), the $\rho$-parts will be right-to-left, and we will add a special left-to-right additional run $\tau$ at the end to come back to the right of the word. Formally, every right-to-right run can be written as
	$\rho_1 \lambda_1 \rho_2 \lambda_2 \cdots \lambda_{m-1} \rho_m \tau$
	where we have a sequence of positions $1\leq i_{m}<\cdots<i_{1}<i_{0}=n+1$ such that
	\begin{itemize}
		\item for all $j\in \{1, \ldots, m\}$, $\rho_j$
		is a left-to-right run over $w_{i_j+1}\cdots w_{i_{j-1}-1}$ with only $\rightarrow$-transitions: notice that this run can be empty if $i_{j-1}=i_j + 1$;
		\item for all $j\in \{1, \ldots, m-1\}$, $\lambda_j$ is a left-to-left run over 
		$w_{i_j}\cdots w_{n}$ that starts with a $\rightarrow$-transition;
		\item $\tau$ is a left-to-right run over $w_{i_m}\cdots w_n$.
	\end{itemize}
	
	We exemplify the decomposition on the right of Figure~\ref{fig:abstract-decomposition}. Once again, the automaton $\A_{(p,q)}$ is transformed into a \NWA where the root automaton is a sweeping automaton that emulates the black $\rho$-parts, interleaved with some new $\leftarrow$-transitions with a weight being a \NWA that computes the $\lambda$-subruns as well as the $\tau$ one. We once again treat these \NWA recursively similarly as before (new cases occur in terms of directions).
	
	This recursive decomposition of the runs, and thus the associated construction of sweeping automata, can be terminated after a bounded number of iterations.\dn{Again I suppose we ask the readers to assume that the head indeed visits some position in the word increasingly many times with each recursive step}Indeed, in all simple runs of $\A$, no more than $|Q|$ configurations are visited for a particular position of the word. Since each recursive step in the decomposition consumes each position in the black runs, this implies that after $|Q|$ steps, there are no remaining blue subruns to consider. At level $|Q|$ of nesting, we thus do not allow anymore the addition of new transitions that would simulate further blue $\lambda$-subruns. The previous argument is the core of the correctness proof showing that the sweeping automaton produced is equivalent to $\A$. 
	
	In case $\A$ is an $\NWAr r$, we use the black $\rho$-subruns to compute the children automata of $\A$ with nested calls. In contrast the added transitions that are supposed to launch the emulation of the blue $\lambda$-subruns call another sweeping automaton below.

\begin{figure}[tbp]
		\centering
		\begin{tikzpicture}[node distance=3cm]
			\node[state, initial left] (init) {$\overrightarrow \iota$};
			\node[state, right of=init, node distance=2cm] (p) {$\overrightarrow p$};
			\node[state, accepting right, accepting by arrow, right of=p, node distance=2cm] (fin) {$\overleftarrow{\kappa}$};
			\node[left of=init, node distance=1.5cm] () {$\overline \A$}; 
			
			\draw[->] (init) edge node[above] {$\lmark, \one, \rightarrow$} (p)
			(p) edge [loop above] node[above]{$a,f,\rightarrow$} (p)
			(p) edge [loop below] node[below]{$b,\overline \A^1_{\overleftarrow p,\overrightarrow p},\rightarrow$} (p)
			(p) edge node[above] {$\rmark,\one,\leftarrow$} (fin);
			
			\begin{scope}[xshift=7.6cm]
				\node (){\includegraphics[scale=.37]{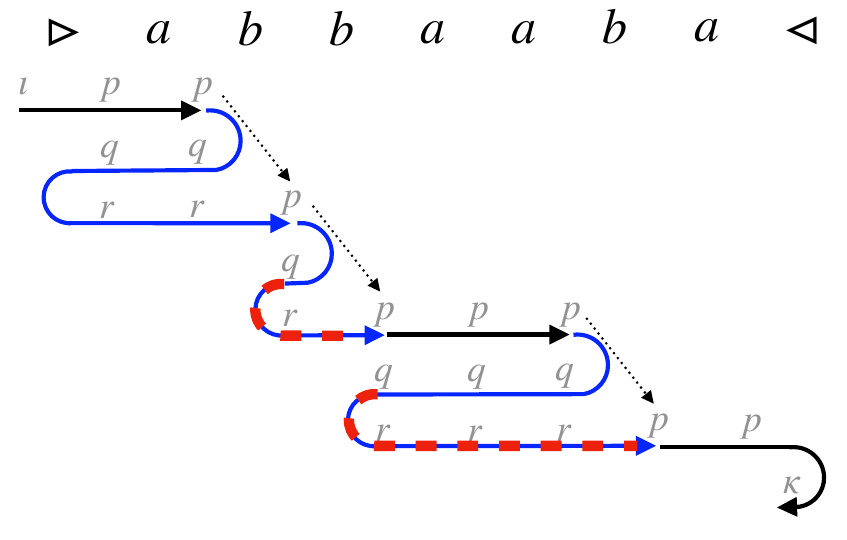}};
			\end{scope}
			
			\begin{scope}[yshift=-3cm,color=blue]
				\node[state,initial] (p) {$\overleftarrow p$} node[left of=p, node distance=1.3cm] () {$\overline \A^1_{\overleftarrow p,\overrightarrow p}$}; 
				\node[state] (q) [right of=p] {$\overleftarrow q$};
				\node[state] (r) [right of=q] {$\overrightarrow r$};
				\node[state,accepting by arrow] (p') [right of=r] {$\overrightarrow p$};
				
				\draw[->] (p) edge [] node [align=center] {$(b,1), \one, \leftarrow$} (q);
				\draw[->] (q) edge [loop above] node [align=center, above] {$(a,0), g, \leftarrow$} (q);
				\draw[->] (q) edge node [align=center, above] {$(\lmark,0), \one, \rightarrow$} (r);
				\draw[->] (q) edge[bend right=20] node [align=center, below] {$(b,0), \overline \A^2_{\overrightarrow q,\overrightarrow p}, \leftarrow$} (p');
				
				\draw[->] (r) edge [loop above] node [above] {$(a,0), \one, \rightarrow$} (r);
				
				\draw[->] (r) edge node [align=center, above] {$(b,1), \one, \rightarrow$} (p');
			\end{scope}
			
			\begin{scope}[yshift=-5cm,color=red]
				\node[state,initial] (q) {$\overrightarrow q$} node [left of=q,node distance=1.3cm] {$\overline \A^2_{\overrightarrow q,\overrightarrow p}$};  
				\node[state] (r) [right of=q] {$\overrightarrow r$};
				\node[state,accepting by arrow] (p) [right of=r] {$\overrightarrow p$};
				
				\draw[->] (q) edge node [align=center, above] {$(b,0,1), \one, \rightarrow$} (r);
				\draw[->] (r) edge [loop above] node [above] {$(a,0,0), \one, \rightarrow$} (r);
				\draw[->] (r) edge node [align=center, above] {$(b,1,0), \one, \rightarrow$} (p);
			\end{scope}
		\end{tikzpicture}
		\caption{$\swNWAr 2$ $\overline \A$ obtained by our construction, starting from the automaton of Example~\ref{ex:running}, and the decomposition of a run of this automaton showing which sweeping automaton computes each subrun.\label{fig:construction}}
	\end{figure}
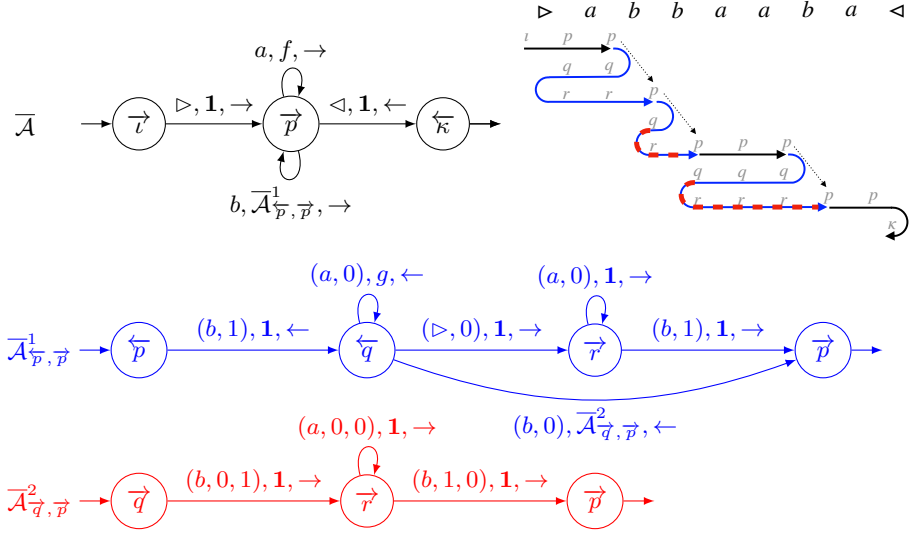

	\begin{example}
		We apply the construction on the $\NWAr 0$ of Example~\ref{ex:running}. This will produce the $\swNWAr 2$ $\overline \A$ in Figure~\ref{fig:construction}. We also depict the actual decomposition of a run over the word $\lmark abbaaba\rmark$. The black subruns are the $\rho$-parts that are computed by the sweeping root automaton (it is sweeping, and not one way, just because of the final transition). The automaton $\overline\A^1_{\overleftarrow{p},\overrightarrow{p}}$ is in charge of computing the blue $\lambda$-subruns. Notice that the subscript tells the automaton that it should start (at the marked position, which is checked by the first transition of the automaton) in state $p$ going left, and should stop (once again at the marked position) in state $p$ going right. There are two cases. For the leftmost $\lambda$-subrun, the sweeping automaton can entirely compute it. For the other ones, it cannot since there is a change of direction in the middle of the word. The red dotted part is thus the $\tau$-final piece of the decomposition in the second step of the recursion, that is taken care of by automaton $\overline \A^2_{\overrightarrow q,\overrightarrow p}$. 
	\end{example}
	
	The above construction preserves the ambiguity of the automata. However, we are not able to directly show that it preserves aperiodicity. We must encode more information in the state space of the various sweeping automata in order to allow for the proof of aperiodicity. In particular, we encode some pieces of information on the behaviours allowed in the current position, allowing us to better understand the structure of the transition monoid of the built automaton. The full construction and proof is given in Appendix \ref{appx:2way}, which concludes the proof of the last implication of Theorem~\ref{thm:contribution-2way}. 
	
	\section{Conclusion}
	
	We have extended the results of Droste and Gastin \cite{droste2019aperiodic} relating restricted weighted first-order logic and aperiodic weighted automata with some restrictions about ambiguity. We thus have closed open questions raised by them, introducing an abstract semantics for a full fragment of weighted first-order logic, and showing the equivalence between this logic and aperiodic nested weighted automaton with linear or polynomial ambiguity. 
	
	We have only studied linear and polynomial ambiguity, contrary to Droste and Gastin that have also characterised finitely-ambiguous and unambiguous aperiodic weighted automata with fragments of the logic. We leave as future work similar study for nested weighted automata, but we do hope that similar restrictions may apply also in our more general case. 
	
	However, dropping the condition on polynomial ambiguity would certainly lead to a logical fragment beyond weighted first-order logic. In particular, the main difficulty is that the logic is not easily able to check the simplicity condition of the accepting runs in this case.
	
	Having introduced two-way navigations (and also nesting) makes possible to ask similar questions to other input structures than words, like finite ranked or unranked trees \cite{DroVog11}, nested words \cite{DroPib12,BolGas13}, or even graphs \cite{BGMZ14a}. Nested weighted automata and weighted logics have already been studied in this setting, without any characterisation of the power of first-order fragments. 
	
	\bibliographystyle{splncs04}

\newpage	
	\appendix
	\section{Example \ref{ex:running}}\label{appx:ex-14}
	
	In this section, we prove that in Example~\ref{ex:running}, there is no equivalent $\swNWAr{0}$ for~$\A_{ex}$.
	
	For the sake of contradiction, suppose there exists a $0\text{-}\swnwA$ $\mathcal{A}_{ex}'$
	such that $\sem{\A_{ex}'}=\sem{\mathcal{A}_{ex}}$. Given a run
	$\rho$ over an input word $\word{u}$, let $\pi_{\rho}(ij)$ be the tuple $\tup{(p_{i_{k}},p_{j_{k}})}$ of runs that occur in $\rho$ when $\A_{ex}'$ reads the positions $i$ and $j$. Note that the length of the tuple $\pi_{\rho}(i)$ is the same for all $i$,  because sweeping automata make full sweeps of the word in every direction. Denote by $\pi_{\rho}(\vartriangleright)$ and $\pi_{\rho}(\vartriangleleft)$ denote the same for the positions that are labelled with the left and right markers resp. Then, the run $\rho$ can be uniquely represented by $\pi_{\rho}(\word{12\cdots|u|})$.
	
	Consider the word $v=(a^{m}b)^{n}$ where $m>1$ and $n$ is large enough so that there exist two positions labelled with a $b$, say $i<j$ such that the corresponding target states of every transition in $\pi_{\rho}(i)$ and $\pi_{\rho}(j)$ are equal.We then show, through a case-by-case analysis, that the existence of the run $\rho$ leads to a contradiction.
	
	\begin{itemize}
		\item Case 0: For every $k$, $\mathsf{wt}(\pi_{\rho}((i+1)\ldots j)_{k})=1$\\
		We consider the word $v'=(a^{m}b)^{\frac{i}{m+1}}(a^{m}b)^{n-\frac{j}{m+1}}$
		and the unique run $\rho'$ over $\word{v'}$ represented by $\pi_{\rho}(\word{1\cdots i(j+1)\cdots|v|})$.
		It is easy to see that $\mathsf{wt}(\rho')=\mathsf{wt}(\rho)$ which
		is clearly a contradiction.
		\item Case 1: For some $k$, $\mathsf{wt}(\pi_{\rho}((i+1)\ldots j)_{k})\in f^{+}$
		(or $g^{+}$)\\
		We then consider the word $v'=v(a^{m}b)^{(j-i)(m+1)}$ and the unique
		run $\rho'$ over $\word{v'}$ represented by $\pi_{\rho}(\word{1\cdots i((i+1)\cdots j)^{m+2}(j+1)\cdots|v|})$.
		It is easy to see that $\mathsf{wt}(\rho')\in(f+g)^{*}f^{m+1}(f+g)^{*}$
		(or $(f+g)^{*}g^{m+1}(f+g)^{*}$) which is clearly a contradiction.
		\item Case 2: For some $k$, $\mathsf{wt}(\pi_{\rho}((i+1)\ldots j)_{k})=f^{r_{1}}g^{s_{1}}\cdots f^{r_{l}}g^{s_{l}}$
		such that $0<\sum r_{i},\sum s_{i}$\\
		It must follow that $\sum(r_{i}+s_{i})\leq j-i$. However, when we
		consider the word $v'=v(a^{m}b)^{2\frac{(j-i)}{m+1}}$ and the unique
		run $\rho'$ over $\word{v'}$ represented by $\pi_{\rho}(\word{1\cdots i((i+1)\cdots j)^{3}(j+1)\cdots|v|})$,
		it is easy to see that $\mathsf{wt}(\rho')\in(f+g)^{*}(f^{r_{1}}g^{s_{1}}\cdots f^{r_{l}}g^{s_{l}})^{3}(f+g)^{*}$
		(or $(f+g)^{*}(g^{s_{l}}f^{r_{l}}\cdots g^{s_{1}}f^{r_{1}})^{3}(f+g)^{*}$)
		which implies that $\sum(r_{i}+s_{i})=2m\frac{j-i}{m+1}$. Using the
		previous observation we then have, $2m\frac{j-i}{m+1}\leq j-i$. Hence,
		$m\leq1$. This is a contradiction.
	\end{itemize}
	
	\noindent Hence there, there does not exist any $0\text{-}\swnwA$ with the
	same semantics as~$\mathcal{A}_{ex}$. 

	\section{Proof of Theorem~\ref{thm:contribution-2way} ($2\Rightarrow 5$)}\label{appx:2way}
	
	In this section, we formally prove the direction $2 \Rightarrow 5$ in Theorem~\ref{thm:contribution-2way}. To do so, we first need to devise some notations and semantics for the constructions that shall ensue.
	
	\begin{definition}
		[Demotion]Given an $\NWAr{r}(\K,A\times\{0,1\})$ $\mathcal{A}=\langle Q,\trans,I,F\rangle$,
		its demotion is the $\NWAr{r}(\K,A\times\{0,1\}^2)$ $\dem{}(\mathcal{A})=\tup{Q,\trans',I,F}$
		where, 
		\begin{align*}
			\trans' & =\{(p,(a,\beta,\gamma),\dem{}(\B),d,q)\mid\beta,\gamma\in\{0,1\},(p,(a,\gamma),\B,d,q)\in\trans\}
		\end{align*}
		and $\dem{}(k)=k$ for $k\in\K$. Intuitively, demotion is a way to push an automaton below in the hierarchy of nesting.
	\end{definition}
	
	For notational convenience, in what follows, we shall relax our definition of an $\NWAr{r}(\K,A)$ by dropping the  initial and final states and allowing some states to have dangling edges. The states with incoming dangling edges would be the new initial states and those with outgoing dangling edges the new final states. One can easily revert to the original semantics by adding unique dummy initial and final states that act as the source and destination of all the dangling edges respectively. Moreover, both directions of this translation preserve aperiodicity in the underlying automaton. Consequently, we replace the sets of initial and final states by subsets of $A\times \NWAr{(r-1)}(\K,A\times\{0,1\}) \times \{\leftarrow,\rightarrow\}\times Q$ and $Q\times A\times \NWAr{(r-1)}(\K,A\times\{0,1\}) \times \{\leftarrow,\rightarrow\}$ respectively, to represent these dangling edges.\dn{so classical semantics still makes sense because our semantics can be converted to classical as explained in paper. We then say that we further simplify representation by just dropping the initial and final states in our semantical automaton (the change is purely syntactic). Since the change occurs in "our" semantics, it does not matter whether the labels are markers. Does that clarify it?}
		
	We are now ready to demonstrate our constructions in two logical steps given as definitions. First, we split the automata into several subautomata, keeping the directional movements intact. Then, we explain how to transform the automata into sweeping ones. 
	
We assume, once again without loss of generality and aperiodicity, that the \emph{input} $\NWA$ has dangling edges that make all accepting runs start from the left marker and stop at the right marker. However, this may not be maintained in our construction for all the descendants.\dn{check the sweeping stage of the constructions}
	
\begin{definition}\label{def:first-step-construction}
		Given a polynomially ambiguous $\NWA(\K,A)$ $\mathcal{A}=\langle Q,\trans,I,F\rangle$, we define a $\NWA(\K,A\times \{0,1\}^n)$ $\mathcal{A}_{b}\tup{\bar{v}_{n}}=\tup{Q,\trans_{n}^b,I_{n}^{b},F_{n}^{b}}$, for all $n\in\N$, $b\in \{0, 1\}$, and sequences $\tup{\bar{v}_{n}} = \tup{\theta_{1},\ldots,\theta_{n}}$ with $\theta_{i}\in(\trans_{n-1}^b)^2\cup\trans_{n-1}^b$ (if $n>0$). For $n=0$, $\tup{\bar{v}_0}$ is the empty tuple and $\mathcal{A}_{b}\tup{\bar{v}_{0}}= \A$. For $n>0$, we also let
		\[\trans_{n}^b = \begin{cases}\begin{array}{l}
			\{ (p,(a, 0,\ldots, 0),\dem{n}(\B),d,q) \mid d\in\{\leftarrow,\rightarrow\},
						a\in A, 
						(p,a,\B,d,q)\in\trans\} \\
			 \cup\{(p,(a_{n},1),\dem{}(\B),\leftarrow,q)\mid (p,a_{n},\B,\leftarrow,q)\in\trans_{n-1}\}\\
			 \color{red}{\cup\{(p,(a_{n-1},1,0),\dem{}(\B),\rightarrow,q) \mid (p,(a_{n-1},0),\B,\rightarrow,q)\in\trans_{n-1}\}}\\
		     \cup\{(p,\lmark,k,\rightarrow,q)\mid (p,\lmark,k,\rightarrow,q)\in\trans\}   \qquad \qquad \text{ if $n+b$ is odd} \\[5mm]
		     \{(p,(a, 0, \ldots, 0),\dem{n}(\B),d,q) \mid d\in\{\leftarrow,\rightarrow\},
						a\in A,
						(p,a,\B,d,q)\in\trans\}\\
				\cup\{(p,(a_{n},1),\dem{}(\B),\rightarrow,q)\mid (p,a_{n},\B,\rightarrow,q)\in\trans_{n-1}\}\\
				\color{red}{\cup\{(p,(a_{n-1},1,0),\dem{}(\B),\leftarrow,q)\mid (p,(a_{n-1},0),\B,\leftarrow,q)\in\trans_{n-1}\}}\\
				\cup\{(p,\rmark,k,\leftarrow,q)\mid (p,\rmark,k,\leftarrow,q)\in\trans\} \qquad \qquad \text{ if $n+b$ is even}
		\end{array}\end{cases}\]
		and distinguish four cases for initial and final dangling edges: 
		\sloppy\begin{enumerate}
			\item If $n+b$ is odd and $\theta_n$ is of the form 
			\[\overset{\hookrightarrow}{\theta_{n}}=((p_{n},a_{n},\B_n,{\leftarrow},q_{n}), (p_{n}',a_{n},\B_n',{\rightarrow},q_{n}')) \in(\trans_{n-1}^b)^2\]
			the automaton should capture the right-to-right runs of $\A$ over the fragment of the input word between the latest marked position and the previously marked position that start with the transition $(p_{n},a_{n},\B_n,{\leftarrow},q_{n})$ and end with the transition $(p_{n}',a_{n},\B_n',{\rightarrow},q_{n}')$. We thus
	let
				\begin{align*}
					I_{n}^b & =\{((a_{n},1),\dem{}(\B_n),{\leftarrow},q_{n})\}\\
					F_{n}^b & =\{(p_{n}',(a_{n},1),\dem{}(\B_n'),{\rightarrow})\}
				\end{align*}
			\item If $n+b$ is odd and $\theta_n$ is of the form \[\overset{\leftarrow}{\theta_{n}}=(p_{n},a_{n},\B_n,{\leftarrow},q_{n})\in\trans_{n-1}^b\] supposing that the first component of $\theta_{n-1}$ is of the form\footnote{If $\theta_{n-1}$ is not a pair, we mean simply $\theta_{n-1}$, though this case will never occur. This is because when $\theta_{n-1}$ is a single transition, the corresponding automaton $\A_b\tup{\overline{v}_{n-1}}$ simulates left-to-right (resp. right-to-left) runs, thus whenever these runs change their direction at a position they must come back to the same position resulting in a right-to-right (resp. left-to-left) run over a fragment of the word. In our constructions such behaviour can only generate pairs of transitions for $\theta_{n}$.} \[(p_{n-1},a_{n-1},\B_{n-1},{\leftarrow},q_{n-1})\in\trans_{n-2}^b\] 
the automaton should capture the right-to-left runs of $\A$ over the fragment of the input word between the latest marked position and the previously marked position that start with the transition $(p_{n},a_{n},\B_n,{\leftarrow},q_{n})$ and end with the transition $(p_{n-1},a_{n-1},\B_{n-1},{\leftarrow},q_{n-1})$. 
We thus let
				\begin{align*}
					I_{n}^b & =\{((a_{n},1),\dem{}(\B_n),{\leftarrow},q_{n})\}\\
					F_{n}^b & =\{(p_{n-1},(a_{n-1},1,0),\dem{2}(\B_{n-1}),{\leftarrow})\}
				\end{align*}
			\item\label{item:first-step-3} If $n+b$ is even and $\theta_n$ is of the form \[\overset{\hookleftarrow}{\theta_{n}}=((p_{n},a_{n},\B_n,{\leftarrow},q_{n}), (p_{n}',a_{n},\B_n',{\rightarrow},q_{n}'))\in(\trans_{n-1}^b)^2\] the automaton should capture the left-to-left runs of $\A$ over the fragment of the input word between the latest marked position and the previously marked position that start with the transition $(p_{n}',a_{n},\B_n',{\rightarrow},q_{n}')$ and end with the transition $(p_{n},a_{n},\B_n,{\leftarrow},q_{n})$. We thus let
				\begin{align*}
					I_{n}^b & =\{((a_{n},1),\dem{}(\B_n'),{\rightarrow},q_{n}')\}\\
					F_{n}^b & =\{(p_{n},(a_{n},1),\dem{}(\B_n),{\leftarrow})\}
				\end{align*}
			\item\label{item:first-step-4} If $n+b$ is even and $\theta_n$ is of the form \[\overset{\rightarrow}{\theta_{n}}=(p_{n},a_{n},\B_n,{\rightarrow},q_{n})\in\trans_{n-1}^b\] supposing that the second component of $\theta_{n-1}$ is of the form \[(p_{n-1}',a_{n-1},\B_{n-1}',{\rightarrow},q_{n-1}')\in\trans_{n-2}^b\] the automaton should capture the left-to-right runs in $\A$ over the fragment of the input word between the latest marked position and the previously marked position that start with the transition $(p_{n},a_{n},\B_n,{\rightarrow},q_{n})$ and end with the transition $(p_{n-1}',a_{n-1},\B_{n-1}',{\rightarrow},q_{n-1}')$. 
			 We thus let
				\begin{align*}
					I_{n}^b & =\{((a_{n},1),\dem{}(\B_n),{\rightarrow},q_{n})\}\\
					F_{n}^b & =\{(p_{n-1}',(a_{n-1},1,0),\dem{2}(\B_{n-1}'),{\rightarrow})\}
				\end{align*}
		\end{enumerate}
		For $b=0$ (resp. $b=1$) we assume that $a_{0} = \lmark$ (resp. $a_{0} = \rmark$), and hence, the (red) transitions over marked variants of $a_{0}$ would not exist. In some cases, this may even mean that the set $F_n^b$ is empty.
	\end{definition}

Recall that an aperiodic $\NWA$ is such that for all elements $x$ of its transition monoid, there exists $p>0$ such that $x^{p+1} = x^p$. Since the transition monoid is finite, we can find the smallest $p>0$ such that for all $x$, $x^{p+1}=x^p$. We call such an integer $p$ the \emph{index} of the aperiodic $\NWA$. 
	
	Below we show that our construction in Definition~\ref{def:first-step-construction} inductively preserves the aperiodicity of the automata.
	
	\begin{lemma}\label{lem:2nwA-aperiodic}
		Suppose that $\mathcal{A}$ is aperiodic. Then
		for every $\bar{v}$ and $b\in \{0,1\}$, $\mathcal{A}_{b}\langle\bar{v}\rangle$ is aperiodic.
	\end{lemma}
	
	\begin{proof}
		We shall prove that if, for an arbitrary $\tup{\bar{v}} = \tup{\theta_{1},\ldots,\theta_{n-1}}$
		and $b\in\{0,1\}$, both $\mathcal{A}$ and $\mathcal{A}_{b}\langle\bar{v}\rangle$
		are aperiodic, then $\mathcal{A}_{b'}\langle\bar{v},\theta_{n}\rangle$
		is also aperiodic, for $n\geq1$ and $b'\in\{0,1\}$. Let $\bar{k}$ be
		the index of $\mathcal{A}$. It then suffices
		to prove the following claim:
		
		\sloppy\begin{claim}
			There exists some $k\in\mathbb{N}$ such that for every word $w = w_1\ldots w_{|w|}\in A_{n}^{*}$,
			$\mathsf{bh}_{e}^{\mathcal{A}_{b}\langle\bar{v},\theta_{n}\rangle}(w^{k})=\mathsf{bh}_{e}^{\mathcal{A}_{b}\langle\bar{v},\theta_{n}\rangle}(w^{k+1})$,
			where $e\in\{ll,lr,rl,rr\}$ and $b\in\{0,1\}$. 
		\end{claim}
		
			Suppose $w\in(A\times\{0\}^{n})^{*}$. Let $\pi_{1}^{*}$
			be the lifting to words of the projection function $\pi_{1}$ that projects every tuple on its first component. If $(p,q)\in\bh{\mathcal{A}_{b}\tup{\bar{v},\theta_{n}}}{e}(w^{k})$
			for $e\in\{ll,lr,rl,rr\}$ and $k\geq\bar{k}$ then by definition
			$(p,q)\in\bh{\mathcal{A}}{e}(\pi_{1}^{*}(w)^{k})$. Now,
			since $\mathcal{A}$ is aperiodic with an index $\bar{k}$,
			$(p,q)\in\bh{\mathcal{A}}{e}(\pi_{1}^{*}(w)^{k+1})$. Then by our
			constructions that includes all transitions of $\A$ with all $\{0,1\}^n$ components being $0$, $(p,q)\in\bh{\mathcal{A}\tup{\bar{v},\theta_{n}}}{e}(w^{k+1})$.
			Hence, $\mathcal{A}\tup{\bar{v},\theta_{n}}$ is aperiodic for words\dn{could you elaborate the confusion? I have made some mild changes. Maybe that helps?}
			in $(A\times\{0\}^{n})^{*}$ with an index of $\overline{k}$.\footnote{This means that we consider aperiodicity of the automaton after restricting the inputs to $(A\times\{0\}^n)^*$}
			
			It remains to prove the claim for a word $w\in {A_{n}}^{*}\setminus(A\times\{0\}^{n})^{*}$.
			We shall prove the claim only for the case~\ref{item:first-step-3} of definition \ref{def:first-step-construction}, 
			i.e. \dn{explained further} $e=ll$ since all the runs are left-to-left. Other cases, follow a similar approach.

			Now consider first the case where $1 = i<j$. Then, by the virtue of our constructions, any run of $\mathcal{A}_{b}\tup{\bar{v},\overset{\hookleftarrow}{\theta_{n}}}$
			over the word $w^{k'}$, for any $k'$, never\dn{that is how the constructions are designed. I have added a phrase at the start of sentence}
			escapes the fragment $w_{1}\cdots w_{j}$,
			until the run ends. Considering the input $w^{k'+1}$ still preserves
			the said slice, i.e. if $(p,q)\in\bh{\mathcal{A}_{b}\tup{\bar{v},\overset{\hookleftarrow}{\theta_{n}}}}{ll}(w^{k'})$
			then $(p,q)\in\bh{\mathcal{A}_{b}\tup{\bar{v},\overset{\hookleftarrow}{\theta_{n}}}}{ll}(w^{k'+1})$. In particular, this holds even when $k' = 1$.
			
			Suppose $1\neq i < j$, then any run over the word $w^{k'}$, for any $k'$, which reads the position $i$ can never be a left-to-left run because again, by the virtue of our construction, once a run reads the position $i$, it cannot go to the left of $i$ until the end of the run. Since, $i \neq 1$, the run will never reach the left end again. The same will be true if the run reads another position where the last component of its label is $1$. Since, there do not exist any transitions over labels marked only in the positions $l < n-1$, there does not exist any run that reads such positions. The only remaining possibility is when the run, say $\rho$, is essentially a left-to-left run over the initial fragment of the word in $(A\times\{0\}^n)^*(A\times\{0\}^{n-2}\times\{1\}\times\{0\} + \epsilon)$. This means, the run $\rho$ continues to exist over the word $w^{k'+1}$, i.e. $\bh{\A_b\tup{\overline{v},\overset{\hookleftarrow}{\theta_{n}}}}{}(\rho)\in \bh{\A_b\tup{\overline{v},\overset{\hookleftarrow}{\theta_{n}}}}{ll}(w^{k'+1})$. In particular, this also holds for $k' = 1$.
			
			Suppose then that $j < i$. Then for the reasons stated above, once again, any run $\rho$ over the word $w$, will be a left-to-left run over the initial fragment of the word $w_1\ldots w_{j - 1}$. Since this run also continues to exist over the word $w^{2}$, we must have $\bh{\A_b\tup{\overline{v},\overset{\hookleftarrow}{\theta_{n}}}}{}(\rho)\in \bh{\A_b\tup{\overline{v},\overset{\hookleftarrow}{\theta_{n}}}}{ll}(w^{2})$.
			
			If instead, $i$ is undefined, we may revert back to the previous case stated above, since $i$ does not play any role in the analysis. If $j$ is undefined then we may revert back to one of the cases when $i < j$, since $j$ does not play any role in the analysis.
			
			Once we prove this for the remaining choices of $e$, we shall have the required index of $\A_b\tup{\overline{v},\overset{\hookleftarrow}{\theta_{n}}}$ over $A_n^*$ to be $k=\max(\overline{k},1)$.
			\qed
	\end{proof}
	
	Following the discussion in the proof sketch for Theorem \ref{thm:contribution-2way} in Section \ref{sec:2way2sweeping},
	we now define a construction which takes as input an $\NWA$ and
	yields an equivalent $\NWA$ such that the root automaton
	is an $\swNWA$ and the remaining behaviour of the input is delegated to a child $\NWA$; here we use
	notations from the previous construction. We now assume that the input automaton has classical semantics for its runs in all its layers\dn{I have now added the meaning of classical}. However, after the construction, that may \emph{not} be the case for the automaton.
	
	\begin{definition}
		\label{def:main-construction}For a vector $\langle\bar{v}\rangle=\langle\theta_{1},\ldots,\theta_{n}\rangle$,
		we define the $\NWA(\K,A_n)$ $\sw{\A_b}\tup{\bar{v}} = \tup{\sw{Q^b_n},\sw{\trans_n^b},\sw{I_n^b},\sw{F_n^b}}$ as follows:
		\begin{enumerate}
			{\allowdisplaybreaks
				\item If $n=0$ i.e. $\mathcal{A}\langle\bar{v}\rangle=\mathcal{A}$, we let $b=0$. Then,
				\begin{align*}
					\sw{Q_0^0} & \coloneqq Q\times(A^{*}/\sim_{rr})\cup Q\times\{([w]_{ll},[w]_{lr})\mid w\in A^{*}\}\\
					\sw{\trans_0^0} & \coloneqq\{((p,[w]_{rr}),a,\B,\rightarrow,(q,[wa]_{rr}))\mid(p,a,\B,\rightarrow,q)\in\trans\}\cup\\
					& \hspace{-1cm}\bigcup_{\substack{a\in A\\
							\delta=(p,a,\B_p,\leftarrow,p'),\delta'=(q',a,\B_q,\rightarrow,q)\\
							(p',q')\in\mathsf{bh}_{rr}(\lmark\!w)
						}
					}(\overset{\rightarrow}{\varepsilon_{\delta,\delta'}}=((p,[w]_{rr}),a,\A_0\tup{(\delta,\delta')},\rightarrow,(q,[wa]_{rr})))\cup\\
					& \bigcup_{\substack{\delta=(p,\vartriangleleft,k,\leftarrow,q)\\{}
							[w]_{rr}\in A^{*}/\sim_{rr}
						}
					}\phi_{\delta}=((p,[w]_{rr}),k,\vartriangleleft,\leftarrow,(q,[\epsilon]_{ll},[\epsilon]_{lr}))\\
					& \bigcup_{\substack{a\in A\\
							(p,a,\B,\leftarrow,q)\in\trans\\{}
							[w]_{ll}\in A^{*}/\sim_{ll}\\{}
							[w]_{lr}\in A^{*}/\sim_{lr}
						}
					}((p,[w]_{ll},[w]_{lr}),a,\B,\leftarrow,(q,[aw]_{ll},[aw]_{lr}))\\
					& \hspace{-3cm}\bigcup_{\substack{[w]_{ll}\in A^{*}/\sim_{ll}\\{}
							[w]_{lr}\in A^{*}/\sim_{lr}\\
							a\in A\\
							\delta=(p,a,\B_p,\leftarrow,p'),\delta'=(q',a,\B_q,\rightarrow,q)\in\trans\\
							(q,p)\in\mathsf{bh}_{ll}(w\!\rmark)
						}
					}\overset{\leftarrow}{\epsilon_{\delta,\delta'}}=((q',[w]_{ll},[w]_{lr}),a,,\A_1\tup{(\delta,\delta')},\leftarrow,(p',[aw]_{ll},[aw]_{lr}))\\
					& \bigcup_{\substack{\delta=(p,\lmark,k,\rightarrow,q)\\{}
							[w]_{ll}\in A^{*}/\sim_{ll}\\{}
							[w]_{lr}\in A^{*}/\sim_{lr}
						}
					}\phi_{\delta}=((p,[w]_{ll},[w]_{lr}),\lmark,k,\rightarrow,(q,[\epsilon]_{rr}))\\
					\sw{I^0_0} & :=\bigcup_{(\vartriangleright,k,\rightarrow,p)\in\init}\{(\vartriangleright,k,\rightarrow,(p,[\epsilon]_{rr}))\}\\
					\sw{F^0_0} & :=\bigcup_{\substack{(p,\vartriangleleft,k,\leftarrow)\in\fin\\{}
							[w]_{rr}\in A^{*}/\sim_{rr}
						}
					}\{((p,[w]_{rr}),\vartriangleleft,k,\leftarrow)\}\\
					& \cup\bigcup_{\substack{[w]_{lr}\in A^{*}/\sim_{lr}\\
							f\in F\\
							(p,f)\in\mathsf{bh}_{lr}(w)\\
							\delta=(p,a,\B,\rightarrow,q)\in\trans
						}
					}\{((p,[w]_{ll},[w]_{lr}),a,\A_1\tup{\delta},\leftarrow)\}
				\end{align*}
				
				\item If $n+b$ is odd and $\theta_{n}=\overset{\hookrightarrow}{\theta_{n}}$, then,
				\begin{align*}
					\sw{Q_n^b} & \coloneqq Q\times(A^{*}/\sim_{rr})\cup Q\times\{([w]_{ll},[w]_{lr})\mid w\in A^{*}\}\\
					\sw{\trans_n^b} & :=\bigcup_{\substack{a\in A_{n}\\
							(p,a,\B,\leftarrow,q)\in\trans_n^b\\{}
							[w]_{ll}\in A^{*}/\sim_{ll}\\{}
							[w]_{lr}\in A^{*}/\sim_{lr}
						}
					}((p,[w]_{ll},[w]_{lr}),a,\B,\leftarrow,(q,[\pi_{1}(a)w]_{ll},[\pi_{1}(a)w]_{lr}))\cup\\
					& \hspace{-3cm}\bigcup_{\substack{[w]_{ll}\in A^{*}/\sim_{ll}\\{}
							[w]_{lr}\in A^{*}/\sim_{lr}\\
							a\in A_{n}\\
							\delta=(p,a,\B_p,\leftarrow,p'),\delta'=(q',a,\B_q,\rightarrow,q)\in\trans_n^b\\
							(q,p)\in\mathsf{bh}_{ll}(w)
						}
					}\overset{\leftarrow}{\varepsilon_{\delta,\delta'}}=((q',[w]_{ll},[w]_{lr}),a,\A_b\tup{\bar{v},(\delta,\delta')},\leftarrow,(p',[\pi_{1}(a)w]_{ll},[\pi_{1}(a)w]_{lr}))\\
					& \cup\bigcup_{\substack{[w]_{lr}\in A^{*}/\sim_{lr}\\{}
							[w]_{ll}\in A^{*}/\sim_{ll}\\
							\delta=(p,\vartriangleright,k,\rightarrow,q)\in\trans_n^b
						}
					}\phi_{\delta}=((p,[w]_{ll},[w]_{lr}),\vartriangleright,k,\rightarrow,(q,[\epsilon]_{rr}))\cup\\
					& \{((p,[w]_{rr}),a,\B,\rightarrow,(q,[w\pi_{1}(a)]_{rr}))\mid(p,a,\B,\rightarrow,q)\in\trans_n^b\}\cup\\
					& \hspace{-2cm}\bigcup_{\substack{a\in A\\
							\delta=(p,a,\B_p,\leftarrow,p'),\delta'=(q',a,\B_q,\rightarrow,q)\\
							(p',q')\in\mathsf{bh}_{rr}(\lmark\!w)
						}
					}(\overset{\rightarrow}{\varepsilon_{\delta,\delta'}}=((p,[w]_{rr}),a,\A_{1-b}\tup{\bar{v},(\delta,\delta')},\rightarrow,(q,[w\pi_{1}(a)]_{rr})))\cup\\
					\sw{I_n^b} & :=\{((a_{n},1),\dem{}(\B_n),\leftarrow,(q_{n},[\epsilon]_{ll},[\epsilon]_{lr}))\}\\
					\sw{F_n^b} & :=\bigcup_{\substack{[w]_{lr}\in A^{*}/\sim_{lr}\\
							[w]_{ll}\in A^{*}/\sim_{ll}\\
							(p,p_{n}')\in\mathsf{bh}_{lr}(w)\\
							\delta=(p,(a,0),\B,\rightarrow,q)\in\trans_n^b
						}
					}\varepsilon_{\delta}=((p,[w]_{ll},[w]_{lr}),(a,0),\A_b\tup{\bar{v},\delta},\leftarrow)\\
					&\cup\bigcup_{[w]_{rr}\in A^{*}/\sim_{rr}}((p_{n}',[w]_{rr}),(a_{n},1),\dem{}(\B_n'),\rightarrow)
				\end{align*}
				\item If $n+b$ is odd and $\theta_{n}=\overset{\leftarrow}{\theta_{n}}$, then,
				\begin{align*}
					\sw{Q_n^b} & \coloneqq Q\times(A^{*}/\sim_{ll})\\
					\sw{\trans_n^b} & :=\bigcup_{\substack{a\in A_{n}\\
							(p,a,\B,\leftarrow,q)\in\trans_n^b\\{}
							[w]_{ll}\in A^{*}/\sim_{ll}
						}
					}((p,[w]_{ll}),a,\B,\leftarrow,(q,[\pi_{1}(a)w]_{ll}))\cup\\
					& \hspace{-2cm}\bigcup_{\substack{[w]_{ll}\in A^{*}/\sim_{ll}\\
							a\in A_{n}\\
							\delta=(p,a,\B_p,\leftarrow,p'),\delta'=(q',a,\B_q,\rightarrow,q)\in\trans_n^b\\
							(q,p)\in\mathsf{bh}_{ll}(w)
						}
					}\varepsilon_{\delta,\delta'}=((q',[w]_{ll}),a,\A_b\tup{\bar{v},(\delta,\delta')},\leftarrow,(p',[\pi_{1}(a)w]_{ll}))\\
					\sw{I_n^b} & :=\{((a_{n},1),\dem{}(\B_n),\leftarrow,(q_{n},[\epsilon]_{ll}))\}\\
					\sw{F^b_n} & :=\bigcup_{[w]_{ll}\in A^{*}/\sim_{ll}}((p_{n-1},[w]_{ll}),(a_{n-1},1,0),\dem{2}(\B_{n-1}),\leftarrow)
				\end{align*}
				\item If $0<n<2\lceil\frac{|Q|}{2}\rceil-2$ is such that $n+b$ is even and $\theta_{n}=\overset{\hookleftarrow}{\theta_{n}}$,
				then,
				\begin{align*}
					\sw{Q_n^b} & \coloneqq Q\times(A^{*}/\sim_{ll})\cup Q\times\{([w]_{rr},[w]_{rl})\mid w\in A^{*}\}\\
					\sw{\trans_n^b} & :=\bigcup_{\substack{a\in A_{n}\\
							(p,a,\B,\rightarrow,q)\in\trans_n^b\\{}
							[w]_{rr}\in A^{*}/\sim_{rr}\\{}
							[w]_{rl}\in A^{*}/\sim_{rl}
						}
					}((p,[w]_{rr},[w]_{rl}),a,\B,\rightarrow,(q,[w\pi_{1}(a)]_{rr},[w\pi_{1}(a)]_{rl}))\cup\\
					& \hspace{-3.5cm}\bigcup_{\substack{[w]_{rr}\in A^{*}/\sim_{rr}\\{}
							[w]_{rl}\in A^{*}/\sim_{rl}\\
							a\in A_{n}\\
							\delta=(p,a,\B_p,\leftarrow,p'),\delta'=(q',a,\B_q,\rightarrow,q)\in\trans_n^b\\
							(p',q')\in\mathsf{bh}_{rr}(w)
						}
					}(\overset{\rightarrow}{\varepsilon_{\delta,\delta'}}=((p,[w]_{rr},[w]_{rl}),a,\A_b\tup{\bar{v},(\delta,\delta')},\rightarrow,(q,[w\pi_{1}(a)]_{rr},[w\pi_{1}(a)]_{rl})))\\
					& \cup\bigcup_{\substack{[w]_{rr}\in A^{*}/\sim_{rr}\\{}
							[w]_{rl}\in A^{*}/\sim_{rl}\\
							\delta=(p,\vartriangleleft,k,\leftarrow,q)\in\trans_n^b
						}
					}\phi_{\delta}=((p,[w]_{rr},[w]_{rl}),\vartriangleleft,k,\leftarrow,(q,[\epsilon]_{ll}))\cup\\
					& \bigcup_{\substack{a\in A_{n}\\
							(p,a,\B,\leftarrow,q)\in\trans_n^b\\{}
							[w]_{ll}\in A^{*}/\sim_{ll}
						}
					}((p,[w]_{ll}),a,\B,\leftarrow,(q,[\pi_{1}(a)w]_{ll}))\cup\\
					& \hspace{-2.5cm}\bigcup_{\substack{[w]_{ll}\in A^{*}/\sim_{ll}\\
							a\in A_{n}\\
							\delta=(p,a,B_p,\leftarrow,p'),\delta'=(q',a,B_q,\rightarrow,q)\in\trans_n^b\\
							(q,p)\in\mathsf{bh}_{ll}(w\rmark)
						}
					}\overset{\leftarrow}{\varepsilon_{\delta,\delta'}}=((q',[w]_{ll}),a,\A_{1-b}\tup{\bar{v},(\delta,\delta')},\leftarrow,(p',[\pi_{1}(a)w]_{ll}))\\
					\sw{I_n^b} & :=\{((a_{n},1),\dem{}(\B_n'),\rightarrow,(q'_{n},[\epsilon]_{rr},[\epsilon]_{rl}))\}\\
					\sw{F_n^b} & :=\bigcup_{\substack{[w]_{rr}\in A^{*}/\sim_{rr}\\{}
							[w]_{rl}\in A^{*}/\sim_{rl}\\
							(p,p_{n})\in\mathsf{bh}_{rl}(w)\\
							\delta=(p,(a,0),\B,\leftarrow,q)\in\trans_n^b
						}
					}\varepsilon_{\delta}=((p,[w]_{rr},[w]_{rl}),(a,0),\A_b\tup{\bar{v},\delta},\rightarrow)\\
					& \cup\bigcup_{[w]_{ll}\in A^{*}/\sim_{ll}}\{((p_{n},[w]_{ll}),(a_{n},1),\dem{}(\B_n),\leftarrow)\}
				\end{align*}
				\item If $0<n<2\lceil\frac{|Q|}{2}\rceil-2$ is such that $n+b$ even and $\theta_{n}=\overset{\rightarrow}{\theta_{n}}$, then,
				\begin{align*}
					\sw{Q_n^b} & \coloneqq Q\times(A^{*}/\sim_{rr})\\
					\sw{\trans_n^b} & :=\bigcup_{\substack{[w]_{rr}\in A^{*}/\sim_{rr}\\
							a\in A_{n}\\
							(p,a,\B,\rightarrow,q)\in\trans_n^b
						}
					}((p,[w]_{rr}),a,\B,\rightarrow,(q,[w\pi_{1}(a)]_{rr}))\cup\\
					& \hspace{-2cm}\bigcup_{\substack{[w]_{rr}\in A^{*}/\sim_{rr}\\
							a\in A_{n}\\
							\delta=(p,a,\B_p,\leftarrow,p'),\delta'=(q',a,\B_q,\rightarrow,q)\in\trans_n^b\\
							(p,q)\in\mathsf{bh}_{rr}(w)
						}
					}\varepsilon_{\delta,\delta'}=((p,[w]_{rr}),a,\A_b\tup{\bar{v},(\delta,\delta')},\rightarrow,(q,[w\pi_{1}(a)]_{rr}))\\
					\sw{I_n^b} & :=\{((a_{n},1),\dem{}(\B_n'),\rightarrow,(q'_{n},[\epsilon]_{rr}))\}\\
					\sw{F^b_n} & :=\bigcup_{[w]_{rr}\in A^{*}/\sim_{rr}}((p_{n-1}',[w]_{rr}),(a_{n-1},1,0),\dem{2}(\B_{n-1}'),\rightarrow)
				\end{align*}
				\item If $n=2\lceil\frac{|Q|}{2}\rceil-2$ then,
				\begin{align*}
					\sw{Q_n^b} & \coloneqq Q\\
					\sw{\trans_n^b} & :=\bigcup_{(p,a,\B,\rightarrow,q)\in\trans_n^b}(p,a,\B,\rightarrow,q)\\
					\sw{I_n^b} & := I_{n}^b\\
					\sw{F_n^b} & :=F_{n}^b
				\end{align*}
			}
		\end{enumerate}
	\end{definition}
	
	Notice how in every one of the above reconstructed automata, the root always exhibits sweeping behaviour. In particular, when $\theta_n$ is of the form $\overleftarrow{\theta_n}$ or $\overrightarrow{\theta_n}$, the resulting root automaton is purely one way. When $\theta_n$ is of the form $\overset{\hookrightarrow}{\theta_n}$ or $\overset{\hookleftarrow}{\theta_n}$, in the resulting root automaton, the states can be divided into two clusters separated by transitions reading only markers. Moreover, within each cluster, the transitions are purely one way never reading a marker. Hence, the only way to change directions when the current switches clusters, effectively reading a marker.
	
	So far, we have developed the machinery necessary for our proof. It remains to show how these can be used to prove the equivalence between $\NWA$ and $\swNWA$. However, to do so we must devise further notation to be able to operate on nested runs with convenience. In particular, we would like to efficiently represent the tree-like structure of these runs.
	
	\subsection{Representation of Nested Runs}
	
	We shall represent the runs of $\NWA$ using ordered trees. 
	
	\subsubsection{Ordered Trees}
	
	An \emph{ordered tree} is a tree where the children of every node
	are linearly ordered. Consider a finite set of labels $\Sigma$, then
	the set of all ordered trees with the nodes labelled from $\Sigma$
	is denoted by $\mathcal{T}(\Sigma)$. 
	
	Consider a tree $t\in\mathcal{T}(\Sigma)$, then,
	\begin{itemize}
		\item $t.root\in\Sigma$ denotes the root node of $t$
		\item $t.l\in\mathcal{T}(\Sigma)$ denotes the subtree rooted at the node
		uniquely specified by the path $l$ from the root of $t$ to the root
		of the subtree; $l\in\mathbb{N}^{*}$ is represented as the sequence
		of order indices of every node in the path with respect to its siblings. 
		\item $t.seq\in\Sigma^{+}$ denotes the ordered sequence of children whose
		parent is $t.root$, i.e. if $t.root$ has $n$ children then $t.seq=(t.1.root)\ldots(t.n.root)$;
		$t.first$ and $t.last$ denote the first and last children in $t.seq$
		resp.
	\end{itemize}
	
\begin{figure}
		\centering
		\includegraphics{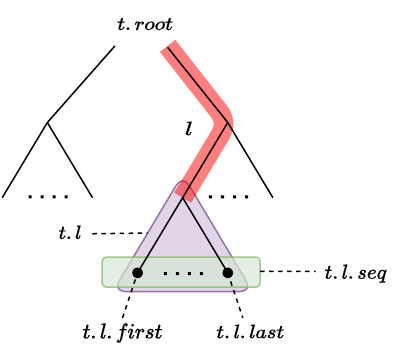}
		\caption{Semantics for an ordered tree $t$\label{fig:ordered-tree}}
	\end{figure}

	Figure \ref{fig:ordered-tree} demonstrates our terminology.
	
	\subsubsection{Runs as Ordered Trees}
	
	Consider a nested run $\rho$; then $\rho$ can be represented as an ordered
	tree where every node denotes a transition in the run. This tree shall
	also be denoted by $\rho$.
	
	A subtree rooted at a node represents the entire behaviour of the
	root transition as executed by the component $\NWA$ that
	corresponds to the weight of that transition. This subtree is denoted
	by $\rho.l$ where $l$ is the unique path from the root of $\rho$
	to the root of the subtree as described for ordered trees.
	
	Two adjacent sibling nodes represent consecutive transitions in the
	behaviour of the component $\NWA$ that corresponds to the
	weight of their parent transition. If the subtree rooted at the parent
	is denoted by $\rho.l$, then $\rho.l.seq$ denotes the ordered sequence
	of all the siblings which are the children of $\rho.l.root$. The
	first and last transition in this sequence can be denoted by $\rho.l.first$
	and $\rho.l.last$; and $\mathsf{bh}(\rho.l)$ denotes the tuple  $\tup{\rho.l.first,\rho.l.last}$.
	
	However, if the automatoon that $\rho$ belongs to, does not have
	a parent automaton, we denote $\rho.root$ by the special symbol $\bullet$.
	
	In addition to the above operations directly inherited from ordered
	trees, we define two more operations on runs,
	
	\subsubsection{Concatenation of runs}
\begin{figure}
		\centering
		\includegraphics{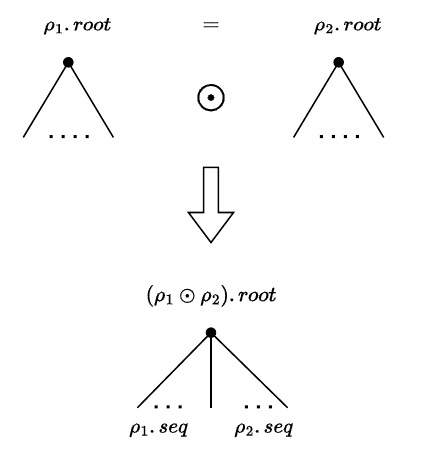}
		\caption{Concatenation of runs\label{fig:concat-runs}}
	\end{figure}

	Given two runs (or subruns) $\rho_{1}$ and $\rho_{2}$, such that
	$\rho_{1}.root=\rho_{2}.root$, their concatenation can be represented
	by the ordered merging of the corresponding trees at the root nodes,
	denoted by $\rho_{1}\odot\rho_{2}$. Binary concatenation can be naturally
	extended to any finite arity. Figure \ref{fig:concat-runs} demonstrates the concatenation of two runs.
	
	It may happen that the concatenation of two runs (or subruns) is not
	a well-formed run (or subrun), i.e. $\rho_{1}.last$ and $\rho_{2}.first$
	may not be synchronised.
	
	We shall alternatively use the terms \emph{runs} and \emph{trees}
	without explicit indication. 
	
	\begin{definition}[Lifting]
		Given a run $\rho$ over an alphabet $A\times\{0,1\}$, it can be
		syntactically \emph{lifted} to obtain the run $\lift{\rho}$ over the alphabet
		$A$ where, $\lift{\rho}.root=\bullet$ and $\lift{\rho}.l.root=(p,a,\leftrightarrow,q)$
		if $\rho.l.root\in\{p\}\times(\{a\}\times\{0,1\})\times\{\leftrightarrow\}\times\{q\}$.
	\end{definition}
	
	\begin{definition}[Flattening]
		Given a word $w=w_{1}\ldots w_{m}$ and an $\NWAr{r}$
		of the form $\sw{\A_b}\tup{\overline{v}}$, consider
		an $r$-nested run $\rho$ over $w$ such that $\rho.seq=\rho_{1}\ldots\rho_{n}$.
		We then recursively define, 
		\[
		\mathsf{flatten}(\rho):=\begin{cases}
			\bigodot_{1\leq i\leq n}\lift{\mathsf{flatten}(\rho.i)} & \!\!\!\!\text{if }\rho\text{ is a run by }\A_{b'}\tup{\overline{v'}}\text{ for some vector }\overline{v'}\\
			\rho' & \!\!\!\!\text{otherwise, if }\rho.root\in(p,a,d,q)_{\mathsf{sw}}
		\end{cases}
		\]
		where $d\in\{\rightarrow,\leftarrow\}$ and $(\cdot)_{\mathsf{sw}}\subseteq\trans_{n}^b\times\sw{\trans_{n}^b}$
		is the relation such that $(p,a,d,q)_{\mathsf{sw}}$
		is the set of transitions in $\sw{\A_b}\tup{\overline{v}}$
		derived from the transition $(p,a,d,q)$ in $\A_b\tup{\overline{v}}$,
		for instance, the transition $((p,[w]_{rr}),a,\rightarrow,(q,[w\pi_{1}(a)]_{rr}))$
		is derived from the transition $(p,a,\rightarrow,q)$. The run $\rho'$
		is then such that $\rho'.root=(p,a,d,q)$ and $\rho'.l=\rho.l$
		for $l\in\mathbb{N}^{+}$.\footnote{Here, $\mathbb{N}^+$ denotes the set of non-empty words over the alphabet $\mathbb{N}$ of natural numbers, \emph{not} the set of positive natural numbers}
	\end{definition}
	
	It is immediate from our definitions that $\mathsf{flatten}(\rho)$
	is a valid run in $\A_b\tup{\overline{v}}$ with equivalent
	semantics to $\rho$. This yields the following result,
	\begin{lemma}\label{lem:right-left}
		Given a word $w=w_{1}\ldots w_{h}\in A^{*}$
		and a vector $\langle\bar{v}\rangle=\langle\theta_{1},\cdots,\theta_{n}\rangle$
		such that $n<2\lceil\frac{|Q|}{2}\rceil-2$, consider a $\NWAr{r}$
		$\A_b\tup{\bar{v}}$. Then $\sem{\sw{\A_b}}\tup{\bar{v}}(w,\sigma)\subseteq\sem{\A_b\tup{\bar{v}}}(w,\sigma)$
		for some $\sigma=\tup{j_1,\ldots,j_n}$\footnote{Here, we use $(w,\sigma)$ to denote the word $((((w,j_1)\ldots),j_{n-1}),j_n)$ from Section \ref{sec:NWA}}.
	\end{lemma}
	
	\sloppy\begin{lemma}\label{lem:left-right}
		Given a word $w=w_{1}\ldots w_{h}\in A^{*}$
		and a vector $\langle\bar{v}\rangle=\langle\theta_{1},\cdots,\theta_{n}\rangle$,
		consider an $\NWAr{r}(\K,A_n)$ $\A_b\tup{\bar{v}}$\footnote{Recall definition from Definition \ref{def:first-step-construction}}
		such that $n<2\lceil\frac{|Q|}{2}\rceil-2$. Then $\sem{\mathcal{A}_{b}\langle\bar{v}\rangle}(w,\sigma)=\sem{\sw{\mathcal{A}_{b}}\langle\bar{v}\rangle}(w,\sigma)$
		for some $\sigma=\tup{j_1,\ldots,j_n}$ and $b\in\{0,1\}$.
	\end{lemma}
	
	\begin{proof}
		We prove for the case when $r=0$; the other cases are syntactically analogous. For the sake of presentation we shall only discuss the relatively non-trivial cases. In particular we assume $n>0$. Moreover, we shall only demonstrate our proof for the case when $n=1$ and $b=1$. The case when $n=1$ and $b=0$ can be symmetrically derived.
		
		It suffices to only show $\sem{\A_0\tup{\theta_1}}\subseteq\sem{\sw{\A_0\tup{\theta_1}}}$. The other direction follows from our definitions as stated in Lemma \ref{lem:right-left}. We use $w_{0}$ and $w_{h+1}$ to denote $\lmark$ and $\rmark$.
		
		Below we cover the different cases for $\theta_1$. In every case, we show how to construct a unique sweeping run $\sw{\rho}$ in $\sw{\A_b\tup{\theta_1}}$ from an arbitrary run $\rho$ in $\A_b$ with equivalent semantics.
		
		\noindent \textit{$(\theta_1 = \overset{\hookleftarrow}{\theta_1})$}
		Consider a left-to-left run $\rho$ over the word $(w_{\sigma_1},1)\ldots\rmark\in (A\times\{0,1\})^*\rmark$. There are two possibilities, (1) $\rho$ visits the marker $\rmark$ or (2) it does not. We further divide our proof into two parts,
		\begin{enumerate}
			\item Suppose $\rho$ visits the right marker. Then we decompose the run as $\overrightarrow{\rho}(p,\rmark,k,\leftarrow,q)\overleftarrow{\rho}$ where $\overrightarrow{\rho}$ is a left-to-right run over $(w_{\sigma_1},1)\ldots(w_n,0)$ and $(p,\rmark,k,\leftarrow,q)\overleftarrow{\rho}$ is a right-to-left run over $(w_{\sigma_1},1)\ldots\rmark$.
			
			Now consider the run $\overrightarrow{\rho}$. It can be further decomposed as $\rho_1\lambda_1\rho_2\lambda_2\ldots\rho_m$ where we have a sequence of positions $\sigma_n-1=i_{0}<i_{1}<\cdots<i_{m-1}<i_{m}=n$ such that
			\begin{itemize}
				\item for all $j\in \{1, \ldots, m\}$, $\rho_j$
				is a run over $(w_{i_{j-1}+1},0)\cdots (w_{i_{j}-1},0)$ with only $\rightarrow$-transitions: notice that this run can be empty if $i_{j}=i_{j-1} + 1$;
				\item for all $j\in \{1, \ldots, m-1\}$, $\lambda_j$ is a right-to-right run over 
				$(w_{\sigma_n},1)\cdots (w_{i_j},0)$ that starts with a $\leftarrow$-transition.
			\end{itemize}
			
			We define $\sw{\rho_j}$ to be the run in $\sw{\A_1\tup{\overset{\hookleftarrow}{\theta_1}}}$ derived from $\rho_j$ by the point-wise substitution of every transition $(p,w_{j'},\B,\rightarrow,q)$ by $((p,[w_{\sigma_n}\ldots w_{j'-1}]_{\sim_{rr}},[w_{\sigma_n}\ldots w_{j'-1}]_{\sim_{rl}}),w_{j'},\B,\rightarrow,(q,[w_{\sigma_n}\ldots w_{j'}]_{\sim rr},[w_{\sigma_n}\ldots w_{j'}]_{\sim rr}))$. The run $\lambda_j$ implies that $(p_j,q_j)\in \bh{\A}{rr}(w_{\sigma_n}\ldots w_{i_j})$ where $(p_j,w_{i_{j}},\B_p,\leftarrow,p_j')=\lambda_{j}.first$
			and $(q_j',w_{i_{j}},\B_q,\rightarrow,q_j)=\lambda^{j}.last$, thus
			we can let $\sw{\lambda_j}=\overrightarrow{\varepsilon_{\lambda_j.first,\lambda_j.last}}$. The run $\overrightarrow{\varepsilon_{\lambda_j.first,\lambda_j.last}}.seq$
			is obtained from $\lambda_j$ by replacing every transition
			reading $w_{i_{j}}$ with the corresponding transition in $\A_1\tup{\overset{\hookleftarrow}{\theta_1},(\lambda_j.first,\lambda_j.last)}$
			reading $(w_{i_{j}},0,1)$ and every other transition reading $a\neq (w_{i_{j}},0)$
			by the corresponding transition that reads $(a,0)$. It follows from the definitions that $\overrightarrow{\varepsilon_{\lambda_j.first,\lambda_j.last}}.seq$ is a run of the automaton $\A_1\tup{\overset{\hookleftarrow}{\theta_1},(\lambda_j.first,\lambda_j.last)}$ over $(\word{w},\tup{\sigma_n,i_{j}})$ with equivalent abstract semantics to $\lambda_j$. Define $\sw{\overrightarrow{\rho}}=\sw{\rho_1}\sw{\lambda_1}\sw{\rho_2}\sw{\lambda_2}\ldots\sw{\rho_m}$.
			
			Next consider the run $\overleftarrow{\rho}$. It can be further decomposed as $\rho_1\lambda_1\rho_2\lambda_2\ldots\rho_m$ where once again we have a sequence of positions $n=i_{0}>i_{1}>\cdots>i_{m-1}>i_{m}=\sigma_n - 1$ such that
			\begin{itemize}
				\item for all $j\in \{1, \ldots, m\}$, $\rho_j$
				is a right-to-left run over $(w_{i_{j}+1},\_)\cdots (w_{i_{j-1}},0)$\footnote{$(a,\_)$ denotes that the position could be either marked or not} with only $\leftarrow$-transitions: notice that this run can be empty if $i_{j}=i_{j-1}$;
				\item for all $j\in \{1, \ldots, m-1\}$, $\lambda_j$ is a left-to-left run over 
				$(w_{i_j},0)\cdots\rmark$ that starts with a $\rightarrow$-transition.
			\end{itemize}
			
			We define $\sw{\rho_j}$ to be the run in $\sw{\A_1\tup{\overset{\hookleftarrow}{\theta_1}}}$ derived from $\rho_j$ by the point-wise substitution of every transition $(p,w_{j'},\B,\leftarrow,q)$ by $((p,[w_{j'+1}\ldots w_n]_{\sim_{ll}}),w_{j'},\B,\leftarrow,(q,[w_{j'}\ldots w_n]_{\sim ll}))$. The run $\lambda_j$ implies that $(q_j,p_j)\in \bh{\A_b\tup{\overset{\hookleftarrow}{\theta_1}}}{ll}(w_{i_j}\cdots\rmark)$ where $(p_j,w_{i_{j}},\B_p,\leftarrow,p_j')=\lambda^{j}.last$
			and $(q_j',w_{i_{j}},\B_q,\rightarrow,q_j)=\lambda^{j}.first$, thus
			we can let $\sw{\lambda_j}=\overleftarrow{\varepsilon_{\lambda_j}.first,\lambda_j.last}$. The run $\overleftarrow{\varepsilon_{\lambda_j.first,\lambda^j.last}}.seq$
			is obtained from $\lambda^j$ by replacing every transition
			reading $w_{i_{j}}$ with the corresponding transition in $\A_1\tup{\overset{\hookleftarrow}{\theta_1},\lambda_j.first,\lambda_j.last}$
			reading $(w_{i_{j}},0,1)$ and every other transition reading $a\neq (w_{i_{j}},0)$
			by the corresponding transition that reads $(a,0)$. It follows from the definitions that $\overleftarrow{\varepsilon_{\lambda_j.first,\lambda_j.last}}.seq$ is a run of the automaton $\A_0\tup{\overset{\hookleftarrow}{\theta_1},(\lambda_j.first,\lambda_j.last)}$ over $(\word{w},\tup{\sigma_n,i_{j}})$ with equivalent abstract semantics to $\lambda_j$. Define $\sw{\overleftarrow{\rho}}=\sw{\rho_1}\sw{\lambda_1}\sw{\rho_2}\sw{\lambda_2}\ldots\sw{\rho_m}$.
			
			Finally, we define the sweeping run $\sw{\rho}\coloneqq\sw{\overrightarrow{\rho}}((p,[w_{\sigma_1}\ldots w_n]_{\sim_{rr}},[w_{\sigma_1}\ldots w_n]_{\sim_{rl}}),\rmark,k,(q,[\epsilon]_{\sim_{ll}})\sw{\overleftarrow{\rho}}$.
			\item Suppose $\rho$ does not visit the right marker. Then we decompose the run as $\overrightarrow{\rho}\tau$ where $\overrightarrow{\rho}$ is a left-to-right run over $(w_{\sigma_1},1)\ldots (w_{i_\tau-1},0)$ for some position $w_{i_\tau}\neq\rmark$ and $\tau$ is a right-to-left run over $(w_{\sigma_1},1)\ldots (w_{i_\tau},0)$.
			Consider the run $\overrightarrow{\rho}$. We can define a run $\sw{\overrightarrow{\rho}}$ just like last time. Let's now turn our attention to $\tau$. The run $\tau$ implies that $(p,p_1)\in\bh{\A}{rl}$ where $(p,w_{i_\tau},\B,\leftarrow,q)=\tau.first$, thus we can let $\sw{\tau}=\varepsilon_{\tau.first}$. The run $\varepsilon_{\tau.first}.seq$ is obtained from $\tau$ by replacing every transition reading $(w_{i_{\tau}},0)$ with the corresponding transition in $\A_0\tup{\overset{\hookleftarrow}{\theta_1},\tau.first}$ reading $(w_{i_{\tau}},0,1)$ and every other transition reading $a\neq(w_{i_\tau},0)$ by the corresponding transition that reads $(a,0)$. It follows that $\varepsilon_{\tau.first}$ is a run of the automaton $\A_1\tup{\overset{\hookleftarrow}{\theta_1},\tau.first}$ over $(\word{w},\tup{\sigma_1,i_\tau})$ with equivalent abstract semantics to $\tau$.
			Finally we define the sweeping run $\sw{\rho}\coloneqq\sw{\overrightarrow{\rho}}\sw{\tau}$.
		\end{enumerate}
		
		\noindent \textit{$(\theta_1 = \overrightarrow{\theta_1})$}
		Consider a left-to-right run $\rho$ over the word $(w_{\sigma_1},1)\ldots (w_{n},0)\in (A\times\{0,1\})^*$. We may decompose it as $\rho_{1}\lambda_{1}\rho_{2}\lambda_{2}\ldots\rho_{m}$ just as shown in the case of $(\theta_{1} = \overset{\hookleftarrow}{\theta_1})(1)$ for $\overrightarrow{\rho}$.
		
		We define $\sw{\rho_j}$ to be the run in $\sw{\A_1\tup{\overrightarrow{\theta_1}}}$ derived from $\rho_j$ by the point-wise substitution of every transition $(p,w_{j'},\B,\rightarrow,q)$ by $((p,[w_{\sigma_n}\ldots w_{j'-1}]_{\sim_{rr}}),w_{j'},\B,\rightarrow,(q,[w_{\sigma_n}\ldots w_{j'}]_{\sim rr})$. The run $\lambda_j$ implies that $(p_j,q_j)\in \bh{\A}{rr}(w_{\sigma_1}\ldots w_{i_j})$ where $(p_j,w_{i_{j}},\B_p,\leftarrow,p_j')=\lambda_{j}.first$
		and $(q_j',w_{i_{j}},\B_q,\rightarrow,q_j)=\lambda^{j}.last$, thus
		we can let $\sw{\lambda_j}=\varepsilon_{\lambda_j.first,\lambda_j.last}$. The run $\varepsilon_{\lambda_j.first,\lambda_j.last}.seq$
		is obtained from $\lambda_j$ by replacing every transition
		reading $w_{i_{j}}$ with the corresponding transition in $\A_1\tup{\overrightarrow{\theta_1},(\lambda_j.first,\lambda_j.last)}$
		reading $(w_{i_{j}},0,1)$ and every other transition reading $a\neq (w_{i_{j}},0)$
		by the corresponding transition that reads $(a,0)$. It follows from the definitions that $\varepsilon_{\lambda_j.first,\lambda_j.last}.seq$ is a run of the automaton $\A_1\tup{\overrightarrow{\theta_1},(\lambda_j.first,\lambda_j.last)}$ over $(\word{w},\tup{\sigma_n,i_{j}})$ with equivalent abstract semantics to $\lambda_j$. Define $\sw{\overrightarrow{\rho}}=\sw{\rho_1}\sw{\lambda_1}\sw{\rho_2}\sw{\lambda_2}\ldots\sw{\rho_m}$.
		
		All the other cases, i.e. $n>1$ follow a similar approach and can be likened to the previous by first replacing the markers by $\sigma_{n-1}$. However, the decomposition of the runs in this case would be slightly different since the runs will never trigger transitions that read markers. Finally it can be checked that $\sem{\sw{\rho}}=\sem{\rho}$.
		
		\qed\end{proof}
	
	We shall now show that our construction preserves aperiodicity,
	\begin{lemma}
		For every $\bar{v}$, if $\mathcal{A}$ is aperiodic then $\sw{\mathcal{A}_{b}}\langle\bar{v}\rangle$
		is also aperiodic, for $n\geq0$ and $b\in\{0,1\}$.
	\end{lemma}
	
	\begin{proof}
		Using Lemma \ref{lem:2nwA-aperiodic}, it suffices to prove that, if $\mathcal{A}_{b}\tup{\bar{v}}$
		is aperiodic then $\mathcal{A}_{b}^{\mathsf{sw}}\tup{\bar{v}}$ must
		also be aperiodic, for every vector $\bar{v}$ and $b\in\{0,1\}$.
		Let the aperiodicity of $\mathcal{A}_{b}\tup{\bar{v}}$ be $\bar{k}$.
		
		We only prove the lemma for the case $\sw{\mathcal{A}_{b}}\tup{\bar{v},\overset{\rightarrow}{\theta_{n}}}$
		when $n=1$ and $b=1$. Other cases follow a similar idea.
		
		Consider the word $w^k=w_1\ldots w_h\in A_{n}^{*}$ and an arbitrary run $\rho$ over this word $w^{k}$, for some $k\geq\bar{k}$,
		with $\bh{\sw{\mathcal{A}_{1}\tup{\overrightarrow{\theta_1}}}}{}(\rho.seq)=((p,[u]_{rr}),(q,[uw^{k}]_{rr}))$. This means
		that the automaton assumes a pre-existing behaviour over the word
		$u$. The run $\rho.seq$ can then be represented as, $\rho_{1}\rho_{2}\ldots\rho_{t}$,
		such that $t$ is odd, where the sub-runs $\rho_{2i-1}$, for $i\leq\lceil\frac{|t|}{2}\rceil$,
		correspond to left-to-right runs over $w_{i}$ without any $\leftarrow$-transitions, in $\mathcal{A}_{1}\tup{\overset{\rightarrow}{\theta_{1}}}$,
		not consisting of transitions of the form $\varepsilon_{\delta,\delta'}$.
		The sub-runs $\rho_{2i}$, for $i\leq\lfloor\frac{|t|}{2}\rfloor$,
		correspond to transitions of the form $\varepsilon_{\delta,\delta'}$,
		which correspond to right-to-right runs over $uw^{\lfloor\frac{i}{|w|}\rfloor}w_{1}\ldots w_{i-\lfloor\frac{i}{|w|}\rfloor}$
		in $\mathcal{A}_{1}\tup{\overset{\rightarrow}{\theta_{1}}}$.
		
		Hence, the run $\mathsf{flatten}(\rho)$ can be re-interpreted as
		a run over $uw^{k}$ in $\mathcal{A}_{1}\tup{\overset{\rightarrow}{\theta_{1}}}$,
		of the form $\rho_{1}\rho_{2}\ldots\rho_{2t'-1}$ which starts from the position $|u|+1$ such that,
		\begin{enumerate}
			\item every run $\rho_{2i-1}$, for $i<t'$, is a left-to-left run over $w^{k}$
			\item every run $\rho_{2i}$, for $i<t'$, is a right-to-right run over $u$
			\item the run $\rho_{2t'-1}$ is a left-to-right run over $w^{k}$
		\end{enumerate}
		Using Lemma \ref{lem:2nwA-aperiodic}, we know that $\bh{}{}(\rho_{j})\in\bh{\mathcal{A}_{1}\tup{\overset{\rightarrow}{\theta_{1}}}}{e}(w^{k+1})$
		for $j\leq2t'-1$ and the apt choice of $e\in\{ll,lr,rl,rr\}$. We can then have a similar run $\rho'$ of the same form such that the behaviours of the new fragments are preserved. It remains to show that the there exists a run $\sw{\rho'}$ in $\sw{\mathcal{A}_{1}}\tup{\overset{\rightarrow}{\theta_{1}}}$ such that $\mathsf{flatten}(\sw{\rho'})=\rho'$. To show this we re-decompose the $\rho'.seq$ as $\rho_1'\lambda_{1}'\ldots\lambda_{m-1}'\rho_{m}'$ where we have a sequence of positions $\sigma_1=i_{0}<i_{1}<\cdots<i_{m-1}<i_{m}=h$ such that,
		\begin{enumerate}
			\item for all $j\in \{1, \ldots, m\}$, $\rho_j$
			is a run over $w_{i_{j-1}+1}\cdots w_{i_{j}-1}$ with only $\rightarrow$-transitions: notice that this run can be empty if $i_{j}=i_{j-1} + 1$;
			\item for all $j\in \{1, \ldots, m-1\}$, $\lambda_j$ is a right-to-right run over 
			$uw_{\sigma_1}\cdots w_{i_j}$ that starts with a $\leftarrow$-transition.
		\end{enumerate}
		Then using a similar approach as in the proof of Lemma \ref{lem:left-right} for the case of $\theta_n = \overrightarrow{\theta_n}$, we can construct the run $\sw{\rho_1'}\sw{\lambda_1'}\sw{\rho_2'}\sw{\lambda_2'}\ldots\sw{\rho_m'}$ with the same behaviour as $\rho$ i.e. 
		$\bh{\sw{\mathcal{A}_{1}\tup{\overrightarrow{\theta_{1}}}}}{}(\sw{\rho_1'}\sw{\lambda_1'}\sw{\rho_2'}\sw{\lambda_2'}\ldots\sw{\rho_m'})=((p,[u]_{rr}),(q,[uw^{k}]_{rr}))$. Since, this is true for arbitrary runs, we thus have $\bh{\mathcal{A}_{1}^{\mathsf{sw}}\tup{\overset{\rightarrow}{\theta_{1}}}}{}(w^{k})\in\bh{\sw{\A_1}\tup{\overset{\rightarrow}{\theta_{1}}}}{}(w^{k+1})$ whence $\sw{\A_1\tup{\overrightarrow{\theta_{1}}}}$ is aperiodic with aperiodicity $\overline{k}$. 
		\qed\end{proof}
	
	\subsubsection{$\mathsf{sw}$-Transformation}
	An \emph{$\mathsf{sw}$-transformation} is a recursive transformation
	that provides a recipe to transform any $\NWA$ to an equivalent
	$\swNWA$. 
	
	Consider a $\NWA(\K,A)$ $\A=\langle Q,\trans,I,F\rangle$
	such that all the component $\NWA$ in $\A$ are
	$\swNWA$. This assumption does not lose any generality as,
	if that is not the case, then we first $\mathsf{sw}$-transform the
	component automata. 
	
	Then the $\mathsf{sw}$-transformation of $\mathcal{A}$, denoted
	by $\mathsf{sw}(\mathcal{A})$, is obtained by iteratively substituting
	every component $\NWA$ of the form $\mathcal{A}_{b}\tup{\overline{v}}$
	by $\sw{\A_b}\tup{\overline{v}}$ until exhaustion.
	In every iteration until the $(2\lceil\frac{|Q|}{2}\rceil-1)^{th}$,
	the substitution results in the generation of children $\NWA$
	of the form $\mathcal{A}_{b'}\tup{\overline{v}}$, $b'\in\{0,1\}$.
	However, in the $(2\lceil\frac{|Q|}{2}\rceil-1)^{th}$ iteration we replace
	the $\NWA$ of the form $\mathcal{A}_{b}\tup{\overline{v}}$ by their left-to-right projections
	where the length of $\overline{v}$ is $2\lceil\frac{|Q|}{2}\rceil-2$.
	By virtue of our construction, this does not generate any more substitutable
	$\NWA$ effectively terminating it. 
	
	Let the $\NWA$ obtained after the $i^{th}$ iteration of
	the $\mathsf{sw}$-transformation of $\mathcal{A}$ be denoted by
	$\mathsf{sw}^{i}(\mathcal{A})$. Note that this means $\mathsf{sw}(\mathcal{A})$
	is \emph{distinct} from $\mathsf{sw}^{1}(\mathcal{A})$. 
	
	While $\mathsf{sw}(\mathcal{A})$ is an $\swNWA$, it remains
	to prove its semantic equivalence with $\mathcal{A}$. We first prove
	the equivalence when $\mathcal{A}$ is a $\NWAr{0}$
	starting with the following lemma which shows that with every nesting of a sweeping child in an $\NWA$, the resulting nested runs visit some position in the input word at least an extra time.
	
	\begin{lemma}
		\label{lem:visit-count}Given a word $w=w_{1}\ldots w_{h}\in A^{+}$
		and a vector $\tup{v}=\tup{\theta_{1},\ldots,\theta_{n}}$, consider an
		$\NWAr{r}$ of the form $\sw{\A_b}\tup{\overline{v}}$
		for $b\in\{0,1\}$. If $\rho$ is a run of $\mathsf{sw}^{1}(\sw{\A_b}\tup{\overline{v}})$
		over $\tup{w,\sigma}$, for some $\sigma=\tup{j_1,\ldots,j_n}$,
		such that for some $\delta_{i}$ in $\rho.seq=\delta_{1}\ldots\delta_{m}$,
		$\rho.i.seq$ is a run of an $\NWAr{(r-1)}$ of the
		form $\sw{\A_{b'}}\tup{\overline{v},\theta_{n+1}}$
		over $\tup{w,\sigma'}$, for some $\sigma'$ and
		$b'\in\{0,1\}$. Moreover, $\rho.i.seq=\delta_{1}'\ldots\delta_{m'}'$
		is such that for some $\delta_{j}'$, $\rho.ij.seq$ is a run of an
		$\NWAr{(r-2)}$ of the form $\sw{\A_{b''}}\tup{\overline{v},\theta_{n+1},\theta_{n+2}}$
		over $\tup{w,\sigma''}$, for some $\sigma''$
		and $b''\in\{0,1\}$. Then all the positions between $\sigma_{n+1}$
		and $\sigma_{n+2}$ (including one of the extremities) which are visited at
		least twice by $\mathsf{flatten}(\rho)$. 
	\end{lemma}
	
	\begin{proof}
		We shall only prove this result for the case when $n+b$, $n+b'$, $n+b''$ are all odd and
		$\theta_n = \overleftarrow{\theta_n}$. The other cases follow a similar approach. 
		
		We first observe that if $\rho.l$ visits a position in the input
		word then $\mathsf{flatten}(\rho.l)$ must do so in the corresponding
		input word as well. 
		
		There exists a fragment of the run $\rho$ which carries the head from
		position $\sigma_n>\sigma_{n+2}$ to position $\sigma_{n+1}$. Without
		loss of generality, we can assume that the fragment is $\delta_{1}\ldots\delta_{i-1}$
		where the last transition in this fragment, $\delta_{i-1}$ moves
		the head to the left to the position $\sigma_{n+})$. Hence, the subrun
		$\mathsf{flatten}(\delta_{1}\ldots\delta_{i-1})$ visits all the positions
		between $\sigma_{n+1}$ and $\sigma_{n+2}$ (including $\sigma_{n+2}$)
		once.
		
		Similarly, there must exist a fragment of the run $\rho.i.seq$ which
		carries the head from position $\sigma_{n+1}$ to position $\sigma_{n+2}$.
		Once again, without loss of generality, we can assume that the fragment
		is $\delta'_{1}\ldots\delta_{j-1}'$ where the first transition, $\delta_{1}'$
		moves the head to the right from $\sigma_{n+1}$ and the last transition
		$\delta_{j-1}'$ moves the head to the right to the position $\sigma_{n+2}$.
		Hence, the subrun $\mathsf{flatten}(\delta'_{1}\ldots\delta_{j-1}')$
		visits all the positions between $\sigma_{n+1}$ and $\sigma_{n+2}$
		(including $\sigma_{n+1}$) once.
		
		Finally, the transition $\delta_{j}'$ and hence the generalised subrun
		$\mathsf{flatten}(\delta_{j}')$ visits the position $\sigma_{n+2}$. 
		
		Thus, in $\mathsf{flatten}(\rho)$, all the subruns shall together
		visit all the positions from $\sigma_{n+1}+1$ to $\sigma_{n+2}$ at
		least twice.
		The other cases might involve considering the possible occurrence of markers.
		\qed\end{proof}
	
	The following theorem is the primary contribution where we prove $(2\Rightarrow 5)$ for the case of $\NWAr{0}$. The general case of $\NWAr{r}$ is just a syntactic extrapolation.
	
	\begin{theorem}
		Consider a polynomially ambiguous, trim and aperiodic $\NWAr{0}(\K,A)$
		$\mathcal{A}_{0}=\langle Q,\trans,I,F\rangle$.
		Then there exists a polynomially ambiguous and aperiodic $\swNWAr{(2\lceil\frac{|Q|}{2}\rceil-2)}$
		with equivalent semantics.
	\end{theorem}
	
	\begin{proof}
		Without loss of generality we shall assume classical semntics for our runs. We shall show that $\mathsf{sw}(\mathcal{A}_{0})$ is the automaton in question, i.e. $\sem{\mathcal{A}_{0}}=\sem{\mathsf{sw}(\mathcal{A}_{0})}$. 
		
		Using Lemma \ref{lem:left-right}, in the $\mathsf{sw}$-transformation
		of $\mathcal{A}_{0}$, every iteration until the $(2\lceil\frac{|Q|}{2}\rceil-2)^{th}$
		preserves the semantic equivalence of the resulting automaton, i.e.
		$\sem{\mathcal{A}_{0}}=\sem{\mathsf{sw}^{i}(\mathcal{A}_{0})}$ for
		$i\leq2\lceil\frac{|Q|}{2}\rceil-2$. Now, in the $(2\lceil\frac{|Q|}{2}\rceil-2)^{th}$
		iteration of our $\mathsf{sw}$-transformation we further generate
		component $\NWA$ of the form $(\mathcal{A}_{0})_{b}\tup{\overline{v}}$
		where the length of $\overline{v}$ is $(2\lceil\frac{|Q|}{2}\rceil-2)$
		and \textbf{$b\in\{0,1\}$}. 
		
		Using Lemma \ref{lem:visit-count}, for any run $\rho$ in $\mathsf{sw}^{(2\lceil\frac{|Q|}{2}\rceil-2)}(\mathcal{A}_{0})$
		of depth\footnote{Recall that every nested run is associated with tree semantics as well} $2\lceil\frac{|Q|}{2}\rceil-2$, there exists some position
		in the input which is visited at least $2\lceil\frac{|Q|}{2}\rceil-1$
		by $\mathsf{flatten}(\rho)$ in $\mathcal{A}_{0}$. Now, any position must
		always be visited odd number of times in any run. This means that
		for any $l\in\mathbb{N}^{2\lceil\frac{|Q|}{2}\rceil-3}$, $\rho.l.seq$
		(if it exists) must be a left-to-right run with only $\rightarrow$-transitions. Otherwise,
		we shall have a run that visits some position more than $|Q|$ times
		and hence contradicts polynomial ambiguity of $\mathcal{A}_{0}$.\footnote{Recall that polynomial ambiguity implies simplicity of runs}
		
		Consequently, for vectors $\overline{v}$ of length $2\lceil\frac{|Q|}{2}\rceil-2$,
		it suffices to define $\sw{(\mathcal{A}_{0})_{b}}\tup{\overline{v}}$
		by simply eliminating all the left transitions in $(\mathcal{A}_{0})_{b}\tup{\overline{v}}$, which aligns with our definitions.
		
		As a result $\sem{\mathsf{sw}^{2\lceil\frac{|Q|}{2}\rceil-1}(\mathcal{A}_{0})}=\sem{\mathcal{A}_{0}}$
		where by definition $\mathsf{sw}^{2\lceil\frac{|Q|}{2}\rceil-1}(\mathcal{A}_{0})$
		is an $\swNWA$. Since $\mathsf{sw}^{2\lceil\frac{|Q|}{2}\rceil-1}(\mathcal{A}_{0})$
		does not generate any component $\NWA$ of the form $(\mathcal{A}_{0})_{b}\tup{\overline{v}}$,
		$\mathsf{sw}^{2\lceil\frac{|Q|}{2}\rceil-1}(\mathcal{A}_{0})=\mathsf{sw}(\mathcal{A}_{0})$.
		\qed
	\end{proof}
	
	\begin{corollary}
		Theorem \ref{thm:contribution-2way} ($2\Rightarrow 5$)
	\end{corollary}
	
\end{document}